\documentclass[usenatbib]{mn2e}
\usepackage{graphicx,times}
\usepackage{bm}
\usepackage{epsf}
\usepackage{amsmath}
\usepackage{amssymb}
\usepackage{color}
\date{\today,~ $ $Revision: 0.9 $ $}
\def\cb{\textcolor{black}}
\def\cs{\textcolor{black}}
\def\la{\langle}
\def\ra{\rangle}
\def\n{\noindent}
\def\be{\begin{equation}}
\def\ee{\end{equation}}
\def\ben{\begin{eqnarray}}
\def\een{\end{eqnarray}}
\def\nn{\nonumber}
\def\oh{\hat\Omega}
\def\myC{{\cal C}}

\def\mI{{\mathcal I}}

\def\rT{{\rm T_0}}

\def\myC{{\cal C}}

\def\bl{{\bf l}}
\def\bx{{\bf x}}
\def\2p{{(2\pi)^2}}

\def\bl{{\bf l}}
\def\be{\begin{equation}}
\def\ee{\end{equation}}
\def\beq{\begin{equation}}
\def\eeq{\end{equation}}
\def\ben{\begin{eqnarray}}
\def\een{\end{eqnarray}}

\def\oh{{\hat\Omega}}
\def\nn{{\nonumber}}

\newcommand{\beqa}{\begin{eqnarray}}
\newcommand{\eeqa}{\end{eqnarray}}

\def\inc{{\int_0^{r_{0}}}}
\def\one{\langle y_{s}^2 \rangle}
\def\two{\langle \kappa(\oh_1) \delta y(\oh_2) \rangle_c}

\begin{document}
\onecolumn
\onecolumn
\title[Cross-correlating $y$ and $\kappa$-maps]
{Cross-correlating Sunyaev-Zel'dovich and Weak Lensing Maps}
\author[Munshi, Joudaki, Coles, Smidt, Kay]
{Dipak Munshi$^{1,2}$, Shahab Joudaki$^{3,4}$, Peter Coles$^{1,2}$, Joseph Smidt$^{4,5}$, Scott T. Kay$^6$\\
$^{1}$School of Physics and Astronomy, Cardiff University, Queen's
Buildings, 5 The Parade, Cardiff, CF24 3AA, UK\\
$^{2}$ Astronomy Centre, School of Mathematical and Physical Sciences, University of Sussex, Brighton BN1 9QH, U.K.\\
$^{3}$ Mail number H30, Swinburne University of Technology, PO Box 218, Hawthorn, Victoria 3122, Australia \\ 
$^{4}$ Department of Physics and Astronomy, University of California, Irvine, CA 92697\\
$^{5}$ Los Alamos National Laboratory, Theoretical Division, P.O. Box 1663, Mail Stop B283, Los Alamos, NM 87545, USA,\\
$^{6}$ Jodrell Bank Center for Astrophysics, Alan Turing Building, The University
of Manchester, M13 9PL, UK}
\maketitle
\begin{abstract}
We present novel statistical tools to cross-correlate frequency cleaned thermal Sunyaev-Zel'dovich (tSZ) maps and
tomographic weak lensing (wl) convergence maps. Moving beyond the lowest order cross-correlation, we introduce a hierarchy of mixed
higher-order statistics, the cumulants and cumulant correlators, to analyse
non-Gaussianity in real space, as well as corresponding polyspectra in the harmonic domain. Using these moments,
we derive analytical expressions for the joint two-point probability distribution function (2PDF)
for smoothed tSZ ($y$) and convergence ($\kappa$) maps. The presence of tomographic information allows us
to study the evolution of higher order {\em mixed} tSZ-weak lensing statistics with redshift.
We express the {\em joint} PDFs $p_{\kappa y}(\kappa,y)$ in terms of individual one-point PDFs [$p_{\kappa}(\kappa)$, $p_y(y)$]
and the relevant bias functions [$b_{\kappa}(\kappa)$, $b_y(y)$]. Analytical results for two different
regimes are presented that correspond to the small and large angular smoothing scales. Results are also obtained
for corresponding {\em hot spots} in the tSZ and convergence maps.
In addition to results based on hierarchical techniques and perturbative
methods, we present results of calculations based on the lognormal approximation. The analytical expressions derived here are
generic and applicable to cross-correlation studies of arbitrary tracers of large scale structure
including e.g. that of tSZ and soft X-ray background. We provide detailed comparison of our analytical
results against state of the art Millennium Gas Simulations with and without non-gravitational effects such as
{\em pre-heating} and {\em cooling} . Comparison of these results with {\em gravity only} simulations,
shows reasonable agreement and can be used to isolate effect of non-gravitational physics from observational data.
\end{abstract}
\section{Introduction}
Free electrons in the Universe can be detected through the inverse
Compton scattering of Cosmic Microwave Background (CMB) photons
\citep{SZ80,SZ72,Bir99,Rep95}. An inhomogeneous distribution of
electrons thus induces a secondary anisotropy in the CMB radiation,
which is proportional to the thermal energy of the electrons
integrated along the line of sight direction. This well-known effect
is called the  thermal Sunyaev-Zel'dovich effect (tSZ) and it is now
routinely being used to image nearby galaxy clusters
\citep{Rs02,Jn05,La06} which the electron temperature can reach the
order of $10 {\rm keV}$. The future seems set for a huge increase in
Sunyaev-Zel'dovich measurements of galaxy clusters; data from
Planck alone has generated 1227 candidates with reasonably high
signal-to-noise ratios \citep{Planck_SZ}. In the intergalactic
medium (IGM), however, the gas is expected to be in mildly overdense
regions and to reach a temperature of only $1{\rm keV}$ or so. While
the tSZ effect from clusters can alter the temperature of the CMB by
an amount of order of ${\rm mK}$ in the Rayleigh-Jeans part of the
spectrum, the contribution from the IGM is expected only to reach
the $\mu{\rm K}$ range. This is below the threshold for detection by
WMAP. \cb{Nevertheless the Planck
Surveyor\footnote{http://www.rssd.esa.int/index.php?project=SP} due
to its wider frequency coverage, higher sensitivity, and improved
resolution~\citep{HBMCD05,Joudaki10}, has recently \cs{published} a near all-sky
map of the y-parameter \citep{Planck_y}; also see \citep{HS14}.}

The unique spectral dependence of the tSZ effect helps in the task
of separating it from other sources of CMB temperature fluctuations
so  various well developed component separation schemes are
available for construction of frequency-cleaned SZ maps
\citep{Leach08,BG99,DCP03}. The construction of such maps will
provide us with a direct opportunity to probe the thermal history of
the Universe, in tandem with other observations:  owing to their
thermal (peculiar motions) ionized electrons scatter CMB photons, an
effect which can be studied using the tSZ effect \citep{RBN02}; the
neutral component of the IGM can be also be probed via observations
of the Lyman-$\alpha$ forest \citep{R98}; X-ray emission due to the
thermal bremmstrahlung that originates from the interaction of
ionized electrons and nuclei can also provide clues to the nature
and evolution of the IGM. These tracers all probe different phases
of the IGM and thus play complementary roles. For example, the X-ray
emission depends on the square of the density, so is most sensitive
to the densest regions in the IGM, primarily in  the local Universe,
while tSZ studies can probe the more distant Universe because
Compton scattering is independent of redshift, as well as being an
unbiased tracer of all electrons, as they all participate in the
scattering, the tSZ effect is weakly dependent on density so probes
the electron density in a wide range of environments.

The modeling of lower order statistics of the tSZ effect can be
carried out using various approaches. In the past
\citep{Sj00,ZP01,KS02,ZS07,AC1,AC2} the modeling has followed the
halo model \citep{CooSeth02}. The statistical distribution of
haloes, specifically their number-density as a function of mass
(i.e. their mass function),  is assumed to be that predicted
provided by the Press-Schechter formalism \citep{PS74} or its
generalisations, and the radial profile of such haloes were assumed
to be that of \cite{NFW96}. The hot gas, assumed to have been heated
by shocks, is taken to be in hydrodynamical equilibrium with the
dark matter distribution. The typical temperature reached in such
systems is sufficient to ionize the hydrogen and helium atoms. These
ingredients are sufficient to model the tSZ effect raising from
collapsed haloes \citep{AC1}; in addition to this analytical
modeling, numerical simulation of the tSZ plays an important role in
our understanding of the physics involved
\cs{\citep{Persi95,Ref00,Sel01,Spr01,White02,Lin04, Z04,
Cao07,Ron07,Hal09,Hal07,dasilva00,Shaw10,Batta12,McCarthy13}}. To extend the halo model to
larger scales, the extended distribution of free electrons is
typically assumed to trace the distribution of dark matter on large
scales where the variations in density are in the linear or
quasi-linear regime. A perturbative approach along these lines has
been developed by several authors over the years
\citep{AC1,AC2,CH00,GS99a,GS99b, Mu11a,Mu11b}.

Ongoing and proposed ground surveys such as SZA\footnote{http://astro.uchicago.edu/sza},
ACT\footnote{http://www.physics.princeton.edu/act}, APEX\footnote{http://bolo.berkeley.edu/apexsz},
SPT\footnote{http://pole.uchicago.edu} and the recently completed all sky
survey Planck has produced
a map of the entire y-sky with a great precision \citep{Planck_y}. The SPT collaboration has already reported the
measurement of the tSZ power spectrum at $\ell \approx 3000$ \cs{\citep{Lu10,Saro13,Hanson13,Holder13,Vieira13,Hou14,Story13,High12}}; the 
ACT collaboration has also reported
analysis on similar scales \cs{\citep{Fw10,Dn10, Shegal11,Hand11,Sher11,Wil13,Dunk13,Calab13}}. It is expected that ongoing surveys will improve these
measurements due to their improved sky coverage as well as wider frequency range.
It is important to realize why
the study of secondaries such as tSZ should be an important aspect of any CMB mission.
In addition to the important physics the secondaries probe,
accurate modeling of the secondary non-Gaussianities is required to
avoid $20\%-30\%$ constraint degradations in
future CMB data-sets such as Planck and CMBPol\footnote{http://cmbpol.uchicago.edu/} \citep{Smidt10}.

While the tSZ surveys described above provide a direct probe of the baryonic Universe, weak lensing
observations \citep{MuPhysRep08} on the other hand can map the dark matter distribution in an unbiased way.
In recent years there has been tremendous progress on the technical front in terms of specification
and control of systematics in weak lensing observables. There are many ongoing and future weak lensing surveys such as
CFHT{\footnote{http://www.cfht.hawai.edu/Sciences/CFHLS/}}
legacy survey, Pan-STARRS{\footnote{http://pan-starrs.ifa.hawai.edu/}}
and the Dark Energy survey (DES){\footnote{https://www.darkenergysurvey.org/}}
and in future, the Large Synoptic Survey Telescope (LSST){\footnote{http://www.lsst.org/llst\_home.shtml}},
Joint Dark Energy Mission (JDEM){\footnote{http://jdem.gsfc.nasa.gov/}} will map the dark matter and dark energy distribution
of the entire sky in unprecedented details.
Among other things, hold great promise in shedding light on
the nature of the dark energy and the origin of neutrino masses \citep{JK11}, where the
weak lensing signals dominate the others considered by e.g. the Dark Energy Task Force \citep{Al11}.
However, the optimism associated with lensing is predicated on overcoming the vast systematic uncertainties in both
measurement and in theory \citep{HS04,MHH05,CH02,HS03,Wh04,Hu06,MTC06}.
The statistics of the weak lensing convergence have been studied in great detail using an extension of perturbation theory \citep{MuJa00,MuJai01,MuVaBa04} and
methods based on the halo model \citep{CH00,TJ02,TJ03}. These studies developed techniques that can be used to predict the lower-order moments (equivalent to the
power spectrum and multi-spectra in the harmonic domain) and the entire PDF
for a given weak lensing survey. The photometric redshifts of source galaxies are useful
for tomographic studies of the dark matter distribution and establish a three-dimensional
picture of their distribution \citep{MKHC11}.

Due to the line of sight integration inherent in the tSZ effect, the redshift evolution is completely lost which degrades the
power of tSZ to distinguish different thermal histories. It is however possible to
recover some of the information lost by cross-correlating the tSZ with external tracers \citep{SZ11,ZP01}.
These tracers could comprise galaxy surveys with spectroscopic redshift information
or dark matter distribution from weak lensing surveys
in tomographic slices. Such cross-correlations has been studied
using two-point statistics. We extend these results to higher order. Higher order statistics such as
cumulants and cumulant correlators \citep{MMC99} can probe higher order cross-correlations
and can in principle provide an independent probe of the bias associated with
the baryonic pressure fluctuations. The tomographic cross-correlation statistics that we develop here can be applied to surveys
with overlapping sky coverage. Many of the weak lensing surveys and the tSZ surveys will
have overlapping sky coverage. For example, DES will have overlap with the SPT sky
coverage and plans to measure photometric redshifts of roughly $10^8$ galaxies up to $z=1.3$.
Tomographic cross-correlation statistics at second or higher-order can provide a
more detailed picture of the evolution of the thermal history of the baryonic gas by mapping the associated
pressure fluctuation. \cb{In a recent paper \citep{WHM13} has cross-correlated the CFHTLenS data with 
Planck tSZ maps. They measure a non-zero correlation between the two maps out to one degree angular separation 
on the sky, with an overall significance of six sigma and use the results 
to conclude a substantial fraction of the "missing" baryons in the universe may reside in a low density 
warm plasma that traces dark matter.} \cs{An internal detection of tSZ and CMB lensing cross-correlation in Planck nominal mission data
has also been reported recently at $6.2$ sigma significance \citep{HS14}.}

This paper is organized as follows. In \textsection\ref{Not} we introduce our notations
for both tSZ and weak-lensing observables. Next we introduce the mixed cumulant correlators and
the cumulants in \textsection\ref{sec:lower}. Two different models are used to model the
underlying dark matter distribution. We employ the \citep{Smith03} prescription, for modelling the evolution of matter
power spectrum and use the hierarchical approach for modelling of the higher order correlation hierarchy.
This allows us to extend the lower order moment results to
arbitrary order and construct the relevant PDF and bias for the smoothed weak lensing 
$\kappa$ field and the $y$ maps in \textsection\ref{sec:pdf}. \cs{ In \textsection\ref{sec:hydro} we present 
a short description of the simulations and \textsection\ref{sec:disc} is reserved for the discussion
of test of analytical results against numerical simulations. Finally, \textsection\ref{sec:conclu} is reserved for concluding remarks.}
A brief review of the
hierarchical ansatz in the quasilinear and non-linear regimes, as well as the
lognormal approach, is provided in the appendix. 

The results presented here are generic and can be extended to
study other secondaries, such as the cross-correlation involving CMB lensing and tSZ maps.
\section{Notations}
\label{Not}
\n
In this section, we introduce our notation for the tSZ effect and weak lensing convergence. We will use the following line element:
\be
ds^2 = -c^2 dt^2 + a^2(r)\left [ dr^2 + d^2_A(r)(\sin^2\theta d\theta^2 + d\phi^2) \right ]
\ee
Here $d_A(r)$ is the comoving angular diameter distance at a (comoving) radial distance $r$.
The angular diameter distance is $d_A(r)= (-{\rm K})^{-1/2}\sinh((-{\rm K})^{1/2}r)$ for negative curvature,
$d_A(r)= ({\rm K})^{-1/2}\sin(({\rm K})^{1/2}r)$ for positive curvature, and for a flat Universe $d_A(r)=r$.
The parameters $H_0$ and $\Omega_0$ decide the constant ${\rm K}$: ${\rm K}=(\Omega_0+\Omega_{\Lambda}-1){\rm H}_0^2$.
The underlying cosmology that we adopt for numerical
study is specified by the following parameter values (to be introduced later):
$\Omega_\Lambda = 0.741,\; h=0.72,\; \Omega_b = 0.044,\; \Omega_{\rm CDM} = 0.215,\;
\Omega_{0} = \Omega_b+\Omega_{\rm CDM},\; n_s = 0.964,\; w_0 = -1,\; w_a = 0,\;
\sigma_8 = 0.803,\; \Omega_\nu = 0$. The comoving radial distance at a redshift $z$ is determined by the
cosmology $(\Omega_0,{\rm H}_0)$
\be
r(z) = \int_0^z {dz \over {\rm H}_0 \sqrt{\Omega_0(1+z)^3 + \Omega_{\rm K} (1+z)^2 + \Omega_{\Lambda} }}
\ee
Throughout, $c$ will denote the speed of light and will be set to unity.
\begin{figure}
\begin{center}
{\epsfxsize=9.5 cm \epsfysize=5.1 cm
{\epsfbox[31 432 584 712]{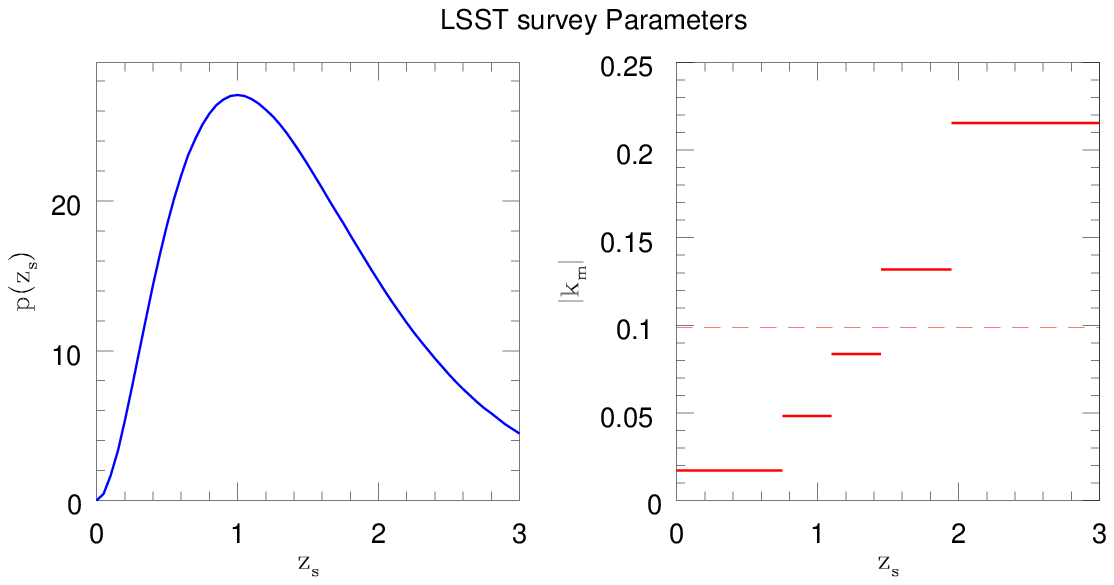}}}
\end{center}
\caption{\cb{The parameter $\kappa_{m}$ for the entire survey and  $\kappa_{m}^{(i)}$ for individual bins (see Eq.(\ref{eq:kappa_min}) for  
definitions), that denotes the minimum value of convergence for the survey or respective bins,} are displayed  as a function of source
redshift for the $\Lambda$CDM cosmology (right panel). These parameters are independent of angular smoothing scale.
For individual bins it is shown using solid lines and for the entire survey it is shown using dashed lines.
The left panel shows the source distribution as a function of redshift as defined in Eq.\ref{eq:source}. \cb{See text for more
details on our choice of tomographic bins.}}
\label{fig:kmin}
\end{figure}
\subsection{Thermal Sunyaev Zel'dovich Effect}
\n
The tSZ effect contributes to the CMB temperature fluctuation and is typically expressed as
$\delta_T(\nu,\oh) = {\Delta T(\oh)/\rT} = g(x_{\nu})y(\oh)$.
In this expression, $g(x_{\nu})$ corresponds to the spectral dependence and $y(\oh)$ encodes
the angular dependence; $x_{\nu}$ represents the dimensionless frequency and $\oh=(\theta,\phi)$ corresponds to a
unit vector that signifies pixel positions. A subscript $s$ will be used to denote the smoothed maps, e.g. $y(\oh)$.
In the non-relativistic limit $g(x_{\nu})$ takes the following form:
\be
g(x_{\nu}) = x_{\nu}\coth \left ({x_{\nu} \over 2} \right ) -4 = \left ( x_{\nu} {e^{x_{\nu}} +1 \over e^{x_{\nu}} -1} - 4\right ); \quad\quad  x_{\nu} = {h\nu \over k_B\rT}= {\nu \over 56.84 {\rm GHz}} = {5.28 {\rm mm} \over \lambda};
\label{eq:def_g}
\ee
\begin{figure}
\begin{center}
{\epsfxsize=9.5 cm \epsfysize=5.1 cm
{\epsfbox[31 432 584 712]{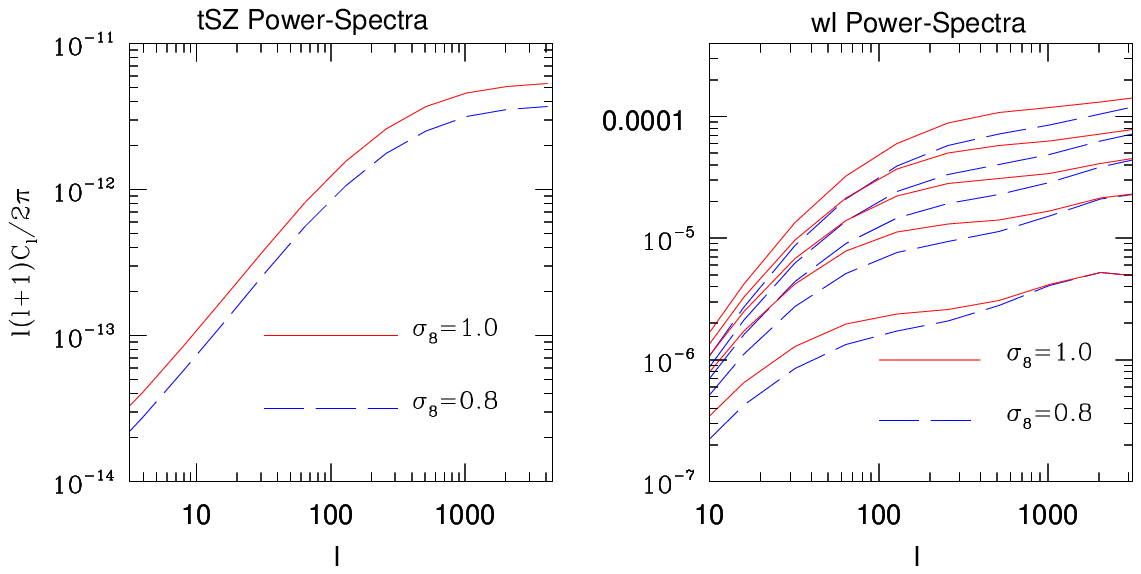}}}
\end{center}
\caption{The left panel shows tSZ power spectrum as a function of the harmonics $\ell$.
The right panel shows redshift-resolved tomographic weak lensing and tSZ cross-spectra, defined in Eq.(\ref{eq:cross_spec}),
as function of $\ell$.
The background cosmology is assumed $\Lambda$CDM. The solid lines in each panel
correspond to $\sigma_8=1$ and the dashed lines correspond to $\sigma_8=0.8$.}
\label{fig:ps}
\end{figure}
Here $k_B$ and $h$ are the Boltzmann and Planck constants respectively; $\nu$ denotes the
frequency of the photon and $\rT=2.726 \; {\rm K}$ is the temperature of the CMB sky. The tSZ effect shows as CMB temperature
decrement at $\nu \ll 218{\;\rm GHz}$ and
as an increment at $\nu \gg 218 {\;\rm GHz}$, with a null at $\nu=218{\;\rm GHz}$. In the Rayleigh Jeans
limit, characterized by $x \ll 1$, $g(x) \approx -2$ is roughly independent of frequency and
in the other limiting situation $x_{\nu} \gg 1$,  $g(x_{\nu}) \approx (x_{\nu}-4)$.
The key information regarding thermal history of the Universe is encoded in $y(\oh)$ maps that
are extracted from frequency maps obtained through multifrequency CMB observations. The $y$ maps
are opacity weighted integrated pressure fluctuations along the line of sight.
\be
y(\oh)\equiv \int ds\; {n_e \sigma_{\rm T}} {k_B T_e \over m_e c^2}= {\sigma_{\rm T} \over m_e c^2} \int_0^{r_{0}} dr\; a(r)n_ek_BT_e(\oh,r) =
{\sigma_{\rm T} \over m_e c^2}\int_0^{\eta_{\rm H}} d\eta \; a(\eta)\; \Pi_e(\eta,\oh) = \int_0^{r_{0}} \; dr\; w_{\rm SZ}(r) \pi_e(r).
\label{eq:weight_tsz}
\ee
%
Here $\Pi_e = n_e k_B T_e$, while $m_e$ corresponds to the electron mass, $k_B$ denotes the Boltzmann's constant,
$\sigma_{\rm T}= 6.65 10^{-25} {\rm cm^2}$ represents the Thompson cross-section, $n_e$ denotes
the number-density of electrons, and $T_e$ is the electron temperature. \cb{We denote the
comoving distance to the surface of last scattering by $r_{0}$}. Conformal time is denoted by $d\eta = dt/a(t)$.
The line-of-sight integral depends on the comoving radial coordinate distance $r$ and $a(r)$ is the corresponding scale factor of the Universe.
The weight is defined as $w_{\rm SZ}(r) = \dot \tau(r) = \sigma_T n_e(r)a(r)$, where the dot defines
the derivative with respect to comoving radial distance $r$, and the \cb{dimensionless} pressure fluctuation is defined as $\pi_e = k_B T_e /m_e c^2$.
\cb{Notice that in accordane with \cite{CTH00}, we have defined the dimensionless pressure $\pi_e$ to be independent of number density of electrons. The $n_e(r)$
dependence of $y$ parameter is absorbed in the weight function $w_{\rm SZ}(r)$ defined above.}
We will cross-correlate the comptonization map $y(\oh)$ with tomographic and projected maps from weak lensing
surveys to constrain the thermal history of the Universe and its evolution with redshift. Throughout,
we will consider the Rayleigh-Jeans part of the spectrum $\delta_T = -2y $; for ACT and SPT operating
at $\nu = 150 {\rm GHz}$ from Eq.~(\ref{eq:def_g}) we get $g(x) = -0.95$. Detailed modelling of
the bias is required only for the computation of variance. The variance  $\la\delta y^2(\oh)\ra$
samples the pressure fluctuation power spectrum $P_{\pi\pi}$ and is expressed as:
\be
\la \delta y^2(\oh)\ra_c = \inc d {r}
{\omega_{\rm SZ}^2(r) \over d^2_A(r)} \int {d^2 {\bf l} \over (2
\pi)^2}~ {\rm P}_{\pi\pi} { \left [ {\ell\over d_A(r)}, r \right ]} b^2_\ell(\theta_s).
\label{eq:ps}
\ee
The pressure power spectrum ${\rm P}_{\pi\pi}(k,z)$ at a redshift $z$ is expressed in
terms of the underlying power spectrum ${\rm P}_{\delta\delta}(k,z)$
using a bias $b_{\pi}(k,z)$, i.e. ${\rm P}_{\pi\pi}(k,z)=b^2(k,z){\rm P}_{\delta\delta}(k,z)$. The bias $b_{\pi}(k,z)$ is assumed
to be independent of length scale or equivalently wave number $k$, i.e. $b_{\pi}(k,z)=b_{\pi}(z)$. The redshift dependent
bias can be expressed as $b_{\pi}(z)= b_{\pi}(0)/(1+z)$. Here $b_{\pi}(0)$ can be written as
$b_{\pi}(0)= k_B T_e(0) b_{\delta}/m_e c^2$. Different values of $b_{\delta}$ have been reported,
e.g. \citep{Ref00} found $b_{\delta} \approx 8-9$ and $T_e(0)\approx 0.3-0.4$. On the other hand
\citep{Sel01} found $b_{\delta} \approx 3-4$.
Typical value of $b_{\pi}(0)$ found by \citep{CO99} is $b_{\pi}(0)=0.0039$.
This is a factor of two lower than the value used by \citep{GS99a,GS99b} and \citep{CH00}. A
Gaussian beam $b_{\ell}(\theta_s)$ with FWHM at $\theta_s$ is assumed.
\cb{We use a redshift dependent generic linear bias model in association with hierarchical
ansatz and shown it's predictions are nearly identical to the predictions of lognormal model.
The linear biasing model has also been used in generating semi-analytical simulations \citep{Wh03,SW03} and
have recently been tested rigorously against numerical simulation \citep{Mu11a}. 
Lognormal model has a higher range of validity than perturbative expansion underlying hierarchical
ansatz. For numerical calculation we have used the log-normal model as it is much easier to implement. We have
found that such an approach works very well for gravity only simulations.}

Some comments are in order at this point. The hierarchical model that we use here was previously used by \cite{ZPW02,CTH00,AC1,AC2}. It is known that the $y$-parameter is proportional to the square of the
density contrast $\delta^2$, so e.g. the three-point correlation function of $y$ effectively
samples the six-point correlation function of the underlying density contrast $\delta$.
However, studies by \citep{AC1,AC2} assumed a linear biasing model for the pressure fluctuations,
i.e. in the Fourier domain $\delta_{\pi}(k,z)= \delta(k,z)b_{\pi}(k,z)$. The pressure bias model
assumes the pressure to be a linear convolution over the density field. Finally \cite{AC1,AC2}
further simplified the bias $b_{\pi}(k,z)$ by factorizing it into a redshift dependent and
momentum $k$ dependent part. The spatial dependence was assumed to be constant. This is
the simplification that we will use in our results, which will be useful in computation of correlation hierarchy
to an arbitrary order. The hierarchical ansatz and the resulting scaling functions
were also used by \citep{VJS01,VS99, VS00}. For computation of the number density of collapsed objects
these were then used to obtain the thermal Sunyaev-Zel'dovich (tSZ) and kinetic Sunyaev-Zel'dovich (kSZ)
effect statistics. The results derived here are complementary to calculations based on the halo model and are
applicable to tSZ effect originating from the extragalactic ionized medium.
\subsection{Weak Lensing in Projection and Tomography}
\begin{figure}
\begin{center}
{\epsfxsize=4.5 cm \epsfysize=5.1 cm
{\epsfbox[31 432 290 712]{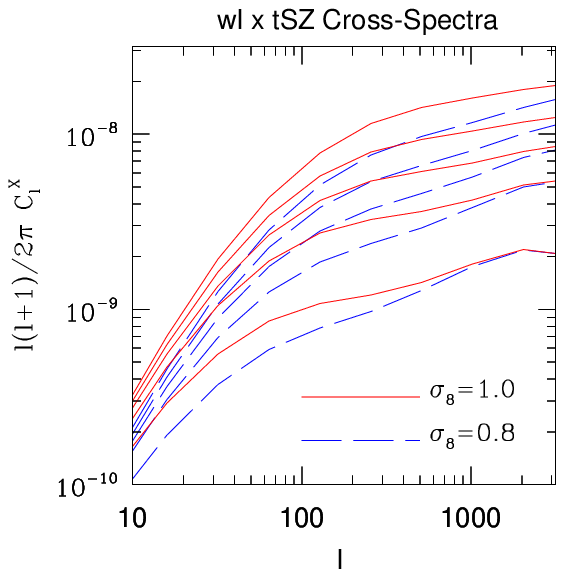}}}
\end{center}
\caption{The redshift-resolved cross spectra $C_\ell^{(i)\kappa,y}$ given in Eq.~(\ref{eq:cross_spec}) for weak lensing tomographic slices and the tSZ
survey are plotted as function of harmonics $\ell$.}
\label{fig:cross}
\end{figure}
Cosmological weak lensing effects are conveniently encoded in the effective convergence field,
which is defined as a weighted projection of the matter density contrast $\delta$.
The statistics of the smoothed weak lensing convergence $\kappa$ are similar to the
projected 3D density contrast $\delta(\bx)$ and can be expressed through a line of sight
integration $\kappa(\oh) = \int_0^{\infty} \; dr ~w_{\rm wl}(r)\delta(r,\oh)$.
Their weight $w_{\rm wl}(r)$ is sensitive to the source distribution;
for 2D surveys without any source redshift information, can be expressed as
\be
\cb{\omega}_{\rm wl}(r) = {3 \over 2}{H_0^2 \over c^2} \Omega_0 a^{-1}(r) d_A(r) \int_{\cb r}^{\cb{\infty}}\; dr_s\; p_s(z_s)\; {dz_s \over dr_s} {d_A(r-r_s) \over d_A(r_s)};
\quad p_s(z_s) = \bar n_g {z_s^2 \over 2 z_0^3} \exp^{-z_s/z_0}.
\label{eq:source}
\ee
Here $p_s(z_s)$ represents the source redshift distribution and $\bar n_g$ denotes surface density of sources. The peak of the distribution is reached at $2z_0$.
We will adopt two different survey configurations. We adopt $z_0=0.5$ for the LSST survey.
For the tomographic calculation we divide the entire source redshift range
in five redshift bins with each tomographic bin containing roughly the same number density of source galaxies.
The redshift bins of the sources are delimited at [0.75, 1.1, 1.45, 1.95, 3.00].
The tomographic convergence maps
$\kappa^{(i)}(\oh) = \int_0^{\cb{\infty}} \; dr ~w^{(i)}_{\rm wl}(r)\delta(r,\oh)$
depend on individual weights:
\be
\cb{\bf \omega}^{(i)}_{\rm wl}(r) = {3 \over 2}{H_0^2 \over c^2} \Omega_0 {1 \over \bar n_i}a^{-1}(r) \; d_A(r)\; \int_{max\{r,r_i\}}^{r_{i+1}}
\;dr\; p_s(r_s)\; {dz_s \over dr_s} {d_A(r-r_s) \over d_{{A}}(r_s)} .
\label{eq:weight_wl}
\ee
We will need the minimum values of the projected convergences for individual tomographic bins $\kappa^{(i)}_{m}$ as well as the
entire survey $\kappa^{}_{m}$ which are defined as follows:  
\be
\kappa^{(i)}_{m} = -\int_0^{\infty} \; dr ~{\cb \omega}^{(i)}_{\rm wl}(r);\quad\quad 
\kappa^{}_{m} = -\int_0^{\infty} \; dr ~{\cb \omega}^{}_{\rm wl}(r).
\label{eq:kappa_min}
\ee
These expressions are derived by setting $\delta=-1$ in the definition of $\kappa^{(i)}$ and $\kappa$ \citep{MuJa00}.
We will next use these expressions to derive the cross-correlations at second- and higher-order both in real space and the harmonic
domain.
%
%
%
%
\section{Mixed Lower order Moments: Cumulants and Cumulant Correlators}
\label{sec:lower}
In this section, we present results in real space and in the harmonic domain that are completely generic
and can provide useful information to study the pressure fluctuations in the baryonic gas. We
apply these results to understand the studies involving tomographic bins from weak lensing surveys.
In case of CMB lensing one of the source planes is identified with the last scattering surface.
The cross-spectra ${\rm P}_{\delta\pi}$ (defined below) can be probed by cross-correlating tomographic weak lensing maps
$\kappa_{(i)}$ and the tSZ $y$ maps of projected pressure fluctuations. These correlations sample the
three dimensional density-pressure cross-spectra ${\rm P}_{\delta\pi}$; using the small angle approximation
\citep{{K98}} one can write:
\begin{equation}
\langle \kappa_{(i)}(\oh_1) \delta y(\oh_2) \rangle_c = \inc d {r}
{{\omega^{(i)}_{\rm wl}(r)\;{\omega_{\rm SZ}(r)}} \over d^2_A(r)} \int {d^2 {\bf l} \over (2
\pi)^2}~\exp ( \;{\cs i}\; {\bf \theta}_{12} \cdot {\bf l} )~ {\rm P}_{\delta\pi} { \left [ {\ell\over d_A(r)}, r \right ]}b_\ell(\theta_s)W_{\rm TH}(\ell\theta_s);
\quad P_{\pi\delta}(k,z) = b_{\pi}(z)P_{\delta}(k,z).
\end{equation}
Here $b_\ell(\theta_s)$ and $W_{\rm TH}(\ell\theta_s)= (2J_1(\ell\theta_s)/\ell\theta_s)$ are Gaussian and Top-hat windows respectively.
The line-of-sight directions $\oh_1$ and $\oh_2$ are separated by an angle $\theta_{12}$. In our notation
$|{\bf l}|=\ell$. The weight function $\omega_{(i)}(r)$ for tomographic convergence and $\omega_{\rm SZ}(r)$
for tSZ surveys are defined in Eq.~(\ref{eq:weight_wl}) and Eq.~(\ref{eq:weight_tsz}) respectively.

To study non-Gaussianity in pressure fluctuation we need to go beyond the power spectrum analysis;
we propose the mixed cumulant correlators for this purpose. Cumulant correlators have the
advantage of being very simple to estimate, and are defined in real space so can be useful for
smaller surveys. Similar results can be obtained for the kurt-spectra,
which we will not consider here.
We will consider a top-hat smoothing $W_{\rm TH}(l\theta_s)$ for the convergence maps and a
Gaussian beam $b_{l}(\theta_s)$ for the $y(\oh)$ maps.
\beqa
\langle \kappa_{(i)}^2 ({\oh_1})\delta y(\oh_2)\rangle_c =&& \inc dr
{[\omega^{(i)}_{\rm wl}]^{\cs 2}(r)\;\omega_{\rm SZ}(r) \over d_A^4(r)} \int {d^2 {\bf l_1} \over (2\pi)^2}
W_{\rm TH}(\ell_1\theta_s) \int {d^2{\bf
l_2}\over (2\pi)^2} W_{\rm TH}(\ell_2\theta_s) \int {d^2 {\bf l_3} \over
(2\pi)^2} b_{l_3}(\theta_s){\rm B}_{\delta\delta\pi} \Big ( {\ell_i\over d_A(r)}, r \Big )_{\sum {\bf l}_i = 0} \\
\langle \kappa_{(i)} ({\oh_1})\;\delta y^2(\oh_2)\rangle_c = && \inc dr
{\omega^{(i)}_{\rm wl}(r)\;\omega^2_{\rm SZ}(r) \over d_A^4(r)} \int {d^2 {\bf l_1} \over (2\pi)^2}
W_{\rm TH}(\ell_1\theta_s) \int {d^2{\bf
l_2}\over (2\pi)^2} b_{\ell_2}(\theta_s) \int {d^2 {\bf l_3} \over
(2\pi)^2} b_{\ell_3}(\theta_s) {\rm B}_{\delta\pi\pi} \Big ( {\ell_i\over d_A(r)}, r \Big )_{\sum {\bf l}_i = 0}
\eeqa
\begin{figure}
\begin{center}
{\epsfxsize=9.5 cm \epsfysize=5.1 cm
{\epsfbox[31 432 584 712]{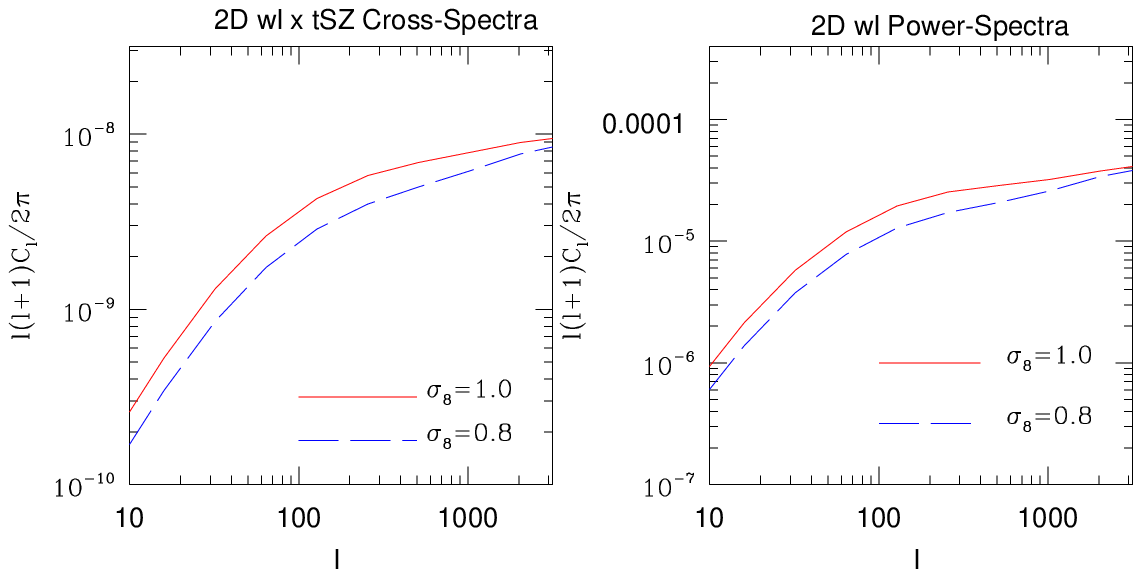}}}
\end{center}
\caption{The projected (or 2D) weak lensing and tSZ cross-spectra  defined in Eq.(\ref{eq:cross_spec}) is plotted as
a function of harmonics $\ell$ (left panel). We also show the projected weak lensing convergence power spectra
as a function of $\ell$ (right panel). Two different values of $\sigma_8 = [0.8,1.0]$ are considered.}
\label{fig:cross2dps}
\end{figure}
\n
Here ${\rm B}_{\delta\delta\pi}$ represents the mixed bispectrum involving three dimensional
density contrast $\delta(\bx)$ and pressure fluctuation $\pi(\bx)$. These {\em mixed} cumulant correlators should
be considered in addition to the pure ones that probe ${\rm B}_{\delta\delta\delta}$ and ${\rm B}_{\pi\pi\pi}$.
These results can trivially be extended to
higher order to compute mixed cumulant correlators. At the level of fourth order they will
probe the mixed trispectra of various types; e.g. we can use $\langle \kappa_{(i)}^2 ({\oh_1}) y^2(\oh_2)\rangle_c$
$\langle \kappa_{(i)}^3 ({\oh_1})y(\oh_2)\rangle_c$ to probe the trispectra  ${\rm T}_{\delta\delta\pi\pi}$,
${\rm T}_{\delta\delta\delta\pi}$ and ${\rm T}_{\delta\pi\pi\pi}$. At these level these statistics are
completely general and they can be estimated in real space simply by cross-correlating smoothed
$y(\theta_s)$ and $\kappa(\theta_s)$ maps raised to suitable powers.
The higher order expressions follow from the same logic.

The derivation so far has been completely generic .
It does {\em not} depend on specific form of the multispectra. Later we will be
use a specific form for the correlation hierarchy which will allow us to include
correlation functions to all order. This is done using a generating function formalism.

The results in the real space are suitable for smaller surveys. However with the increase in
survey size it is pertinent to consider a harmonic space approach as real space measurements
are often correlated. We will introduce power spectra in the harmonic domain that are
Fourier transforms of the corresponding cumulant correlators in real space.
\be
\myC_\ell^{\kappa\kappa,y} = \sum_{\ell_1\ell_2} \sqrt{(2\ell_1+1)(2\ell_2+1)\over 4\pi(2\ell+1)}
\left ( \begin{array}{ c c c }
     \ell_1 & \ell_2 & \ell \\

     0 & 0 & 0
  \end{array} \right) {\cal B}^{\kappa\kappa y}_{l_1l_2l}; \quad
\myC_\ell^{yy,\kappa} = \sum_{\ell_1\ell_2} \sqrt{(2\ell_1+1)(2\ell_2+1)\over 4\pi(2\ell+1)}
\left ( \begin{array}{ c c c }
     \ell_1 & \ell_2 &\ell \\
     0 & 0 & 0
  \end{array} \right) {\cal B}^{yy\kappa}_{\ell_1\ell_2\ell} 
\label{eq:bispec_yk}.
\ee
\cb{Here ${\cal B}^{\kappa\kappa y}_{\ell_1\ell_2\ell_3}$ and ${\cal B}^{\kappa yy}_{\ell_1\ell_2\ell_3}$ denote the angular bispectra. They are related to their
3D counterparts by a line of sight projection e.g.:}
\be
{\cal B}^{\kappa\kappa y}_{\ell_1\ell_2\ell_3} = \int_0^{\infty} dr \left [{w_{\rm wl}^{\cs 2}(r) w_{\rm SZ}(r) \over d^4_A(r) }  \right ] 
{\rm B}_{\delta\delta \pi}\left [{\ell_1\over d_A(r)}, {\ell_2\over d_A(r)},{\ell_3\over d_A(r)};r  \right ].
\ee
Similar expressions can be obtained for individual bins by replacing $w(r)$ with $w_{(i)}(r)$. 
The skew- and kurt-spectra are simple to estimate from real data and the
scatter for such estimates is well understood \citep{MH10,Mn11}. These results can be
used to construct a family of skew-spectra for various choices of tomographic slices.
We also provide the expressions for the auto- and cross-spectra
for tSZ and wl surveys below:
\be
\myC_\ell^{(i)\kappa,y} = \cb{\int_0^{\infty}} 
dr {\omega^{(i)}_{\rm wl}(r)\omega_{\rm SZ}(r) \over d_A^2(r)} {\rm P}_{\delta\pi} \Big ( {\cb{\ell}\over d_A(r)}; r \Big )\\
\label{eq:cross_spec}
\ee
The auto spectra for wl and tSZ can be expressed in terms of ${\rm P}_{\delta\delta}$ and ${\rm P}_{\pi\pi}$ with
suitable changes in the weight functions, i.e. $[\omega^{(i)}_{\rm wl}(r)]^2$ for wl surveys and  $\omega^2_{sz}(r)$ for
tSZ surveys.

\cb{The wl ps for individual bins and the tSZ (y-parameter) ps are displayed in 
Figure \ref{fig:ps}. The projected or 2D tSZ ps is obtained 
by replacing both $\omega^{(i)}_{\rm wl}(r)$ in Eq.(\ref{eq:cross_spec})  by $\omega_{\rm SZ}(r)$.
For computing the wl auto-spectra for individual bins we replace both weight functions by $\omega_{(i)}(r)$.
We have displayed the cross-spectra $\myC_\ell^{(i)\kappa,y}$ defined in Eq.(\ref{eq:cross_spec}) 
in Figure \ref{fig:cross} as a function of the harmonics $\ell$.
We have shown the results for two different $\sigma_8$. Higher redshift bins are more correlated with tSZ. 
The 2D tSZ cross-spectra is obtained  by replacing  $\omega^{(i)}_{\rm wl}(r)$ in Eq.(\ref{eq:cross_spec}) by $\omega_{\rm wl}(r)$. 
The 2D wl ps (right-panel) and the 2D cross-spectra (left-panel) are shown in Figure-\ref{fig:cross2dps}.}
\subsection{Hierarchical {\em Ansatze}}
\label{subsec:hier}
In deriving the above expressions we have not used any specific form
for the
matter correlation hierarchy, however the length scales involved
in small angles are in the highly non-linear regime. Assuming a tree model
for the  matter correlation
hierarchy in the highly non-linear regime one can write the most
general case as 
(White 1979; Peebles 1983; Fry 1984; Bernardeau \& Schaeffer 1992; Szapudi \& Szalay 1993; Bernardeau \& Schaeffer 1999):
\begin{equation}
\xi_N^{\delta}( {\bf r_1}, \dots {\bf r_N} ) = \sum_{\alpha, \rm N-trees}
Q_{N,\alpha} \sum_{\rm labellings} \prod_{\rm edges (i,j)}^{(N-1)}
\xi^{\delta}_2({\bf r_i}, {\bf r_j}).
\end{equation}
It is interesting to note that an exactly similar hierarchy
develops in the quasi-linear regime in the limit of vanishing variance
(Bernardeau 1992), however the hierarchical amplitudes $Q_{N, \alpha}$
become shape-dependent in such a case. In the highly nonlinear
regime there are some indications that these functions become
independent of shape parameters as has been suggested by studies of the
lowest order parameter $Q_3 = Q$ using high resolution numerical
simulations (Sccociamarro et al. 1998). In the Fourier space such an
{\em ansatz} will mean that the whole hierarchy of multi-spectra $B_{\delta},T_{\delta}$
can be written in terms of sum of products of power-spectra $P_\delta$, e.g. in
low orders we can write:
\begin{eqnarray}
B_{\delta}({\bf k}_1, {\bf k}_2, {\bf k}_3)_{\sum k_i = 0} &=& Q ( P_{\delta}({
k_1})P_{\delta}({ k_2}) + P_{\delta}({ k_2})P_{\delta}({ k_3})
+ P_{\delta}({k_3})P_{\delta}({ k_1}) )\delta_D({\bf k_1+k_2+k_3}), \\
T_{\delta}({\bf k}_1, {\bf k}_2, {\bf k}_3, {\bf k}_4)_{\sum k_i = 0} &=& [R_a
P_{\delta}({k_1})P_{\delta}({|\bf k_1 +
k_2|}) P_{\delta}(|{\bf k_1 + k_2 + k_3}|)  + {\rm cyc. perm.} \nn \\
&& + R_b P_{\delta}({k_1})P_{\delta}({ k_2})P_{\delta}({ k_3}) +
{\rm cyc. perm.} ]\delta_D({\bf k_1+k_2+k_3+k_4})
\end{eqnarray}
Different hierarchical models differ in the way they predict the
   amplitudes of different tree topologies. Bernardeau \&
Schaeffer (1992) considered the case where amplitudes in general are
factorizable, at each order one has a new ``star'' amplitude
 and higher order ``snake'' and ``hybrid'' amplitudes are
constructed from lower order ``star'' amplitudes (see Munshi,
Melott \& Coles 1999a,b,c for a detailed description). In models proposed by
Szapudi \& Szalay (1993) it is assumed that all hierarchical amplitudes of a
given order are actually degenerate.
We do not use any of these specific models for clustering and only
assume the hierarchical nature of the higher order correlation functions. In the past, primarily
data from galaxy surveys have been analyzed extensively using these {\em ansatze}. Our main
motivation here is to show that cross-correlation statistics of weak-lensing surveys and the
tSZ surveys can also be analysed using such techniques.
The most general form for the lower order cumulant correlators in the large
separation limit can be expressed as:
\begin{eqnarray}
\langle \kappa^2(\oh_1) \delta y(\oh_2) \rangle_c & = &
2Q_3 \hat{\cal C}_3 [\mI_{\theta_s} \mI_{\theta_{12}}] =
C_{21}^{\eta\eta'}\hat{\cal C}_3 [\mI_{\theta_s} \mI_{\theta_{12}}] \equiv C_{21}^{\kappa y} \langle
\kappa^2 \rangle_c \langle \delta y(\oh_1) \kappa(\oh_2) \rangle_c, \\
\langle \kappa^3(\oh_1) \delta y( \oh_2) \rangle_c & = &
(3R_a + 6 R_b)\hat{\cal C}_4 [\mI_{\theta_s}^2 I_{\theta_{12}}] =
 C_{31}^{\eta\eta'}\hat{\cal C}_4 [\mI_{\theta_s}^2 I_{\theta_{12}}]
 \equiv  C_{31}^{\kappa y} \langle
\kappa^2 \rangle_c^2 \langle \kappa(\oh_1)\delta y(\oh_2) \rangle_c
,\\  \langle \kappa^2(\oh_1) \delta y^2(\oh_2) \rangle_c & =
& 4 R_b\hat{\cal C}_4 [\mI_{\theta_s}^2 \mI_{\theta_{12}}]
= C_{22}^{\eta\eta'}\hat{\cal C}_4 [\mI_{\theta_s} \mI_{\theta_{12}}]
\equiv  C_{22}^{\kappa y} \langle
\kappa^2 \rangle_c\langle \delta y^2 \rangle_c \langle \kappa(\oh_1) \delta y(\oh_2) \rangle_c ,\\
\langle \kappa^4(\oh_1)\delta y(\oh_2)\rangle_c & = &
(24S_a + 36S_b + 4 S_c)\hat{\cal C}_5 [\mI_{\theta_s}^3
\mI_{\theta_{12}}] =
C_{41}^{\eta\eta'} \hat{\cal C}_5 [\mI_{\theta_s}^3 \mI_{\theta_{12}}]
\equiv  C_{41}^{\kappa y} \langle
\kappa^2 \rangle_c^3 \langle \kappa(\oh_1)\delta y(\oh_2)
\rangle_c
,\\ \langle \kappa^3(\oh_1)\delta y^2(\oh_2) \rangle_c & = &
 (12S_a + 6 S_b)\hat{\cal C}_5[I_{\theta_s}^3 I_{\theta_{12}}] =
C_{32}^{\eta\eta'}\hat{\cal C}_5[\mI_{\theta_s}^3 \mI_{\theta_{12}}]
\equiv  C_{32}^{\kappa y} \langle
\kappa^2 \rangle_c^2 \langle
\delta y^2 \rangle_c \langle \kappa(\oh_1)\delta y(\oh_2)
\rangle_c.
\label{eq:cpq_from_tree}
\end{eqnarray}
\begin{figure}
\begin{center}
{\epsfxsize=10. cm \epsfysize=5.1 cm
{\epsfbox[31 432 590 712]{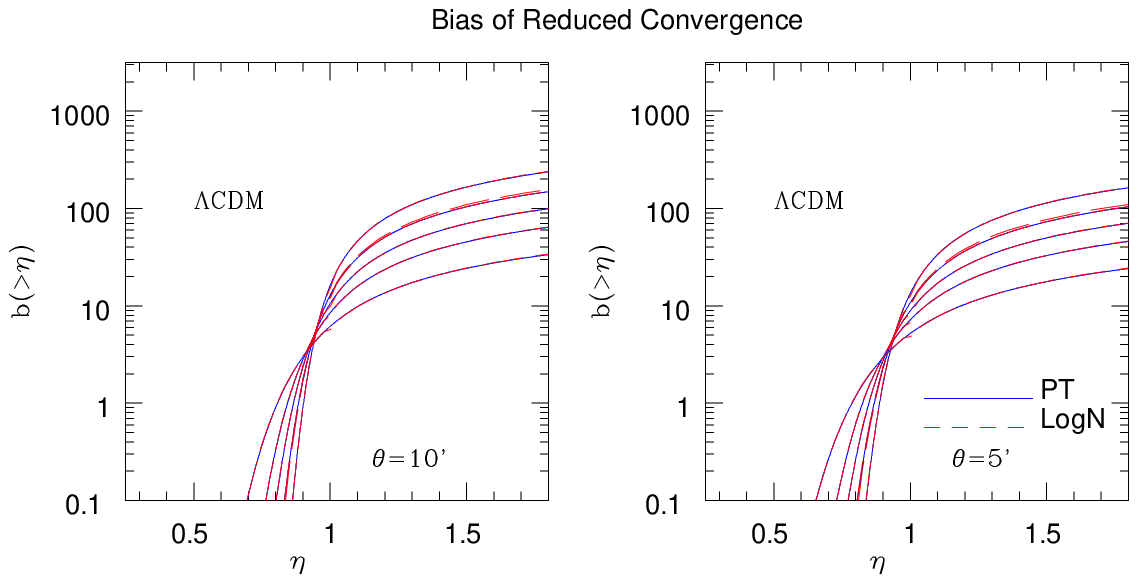}}}
\end{center}
\caption{The bias function $b(\eta)$ \cb{defined in Eq.(\ref{eq:beta})} of the reduced convergence is depicted for two different smoothing angular scales.
The dashed lines show the analytical predictions from the hierarchical ansatz and the solid lines show predictions from
the lognormal distribution. The left panel corresponds to $\theta_s=10'$ and the right panel corresponds
to $\theta_s=5'$. The different curves in each panel correspond to different redshift bins. The lower curves
correspond to shallower (lower redshift) bins and the higher curves correspond to deeper (higher redshift) bins.
The top-most curve corresponds to the projected survey without any tomographic information.}
\label{fig:bias_kappa}
\end{figure}
Here $C_{pq}^{\kappa y}$ denotes the cumulant correlators of the
convergence field and $C_{pq}^{\eta\eta'}$ denotes the cumulant correlators
for the underlying mass distribution. The subscript $c$ denotes
the {\em connected} components of the higher order correlation functions.
The amplitudes $R_a= \nu_2^2$,
$R_b= \nu_3$ and $S_a = \nu_2^3, S_b = \nu_3\nu_2, S_c=\nu_4$ are
expressed in terms of the vertices which can be evaluated
using HEPT (Hyper Extended Perturbation Theory; \cite{Scc98}). The exact expressions
for various $C_{pq}^{\eta\eta'}$ presented in Eq.~(\ref{eq:cpq_from_tree}) are obtained by counting of the relevant tree diagrams.
In the limiting situation when $\oh_1 = \oh_2$ we can recover the
corresponding cumulants. Extending the above results to arbitrary order we can write:
\begin{equation}
 \langle \kappa^p(\oh_1)\delta y^q(\oh_2)\rangle_c  =
 C_{pq}^{\eta}\hat{\cal C}_{p+q}[[{\mathcal I}_{\theta_s}]_{\rm wl}^{(p-1)}[[{\mathcal I}_{\theta_s}]_{\rm sz}^{(q-1)}[{\mI}_{\theta_{12}}]]
= C_{pq}^{\kappa y} [\langle
\kappa^2 \rangle_c^{(p-1)}] [\langle \delta y^2 \rangle_c^{(q-1)}] \langle \kappa(\oh_1)\delta y(\oh_2)
\rangle_c.
\label{eq:mixed_cumu}
\end{equation}
\begin{figure}
\begin{center}
{\epsfxsize=10.5 cm \epsfysize=5.1 cm {\epsfbox[21 432 587 712]{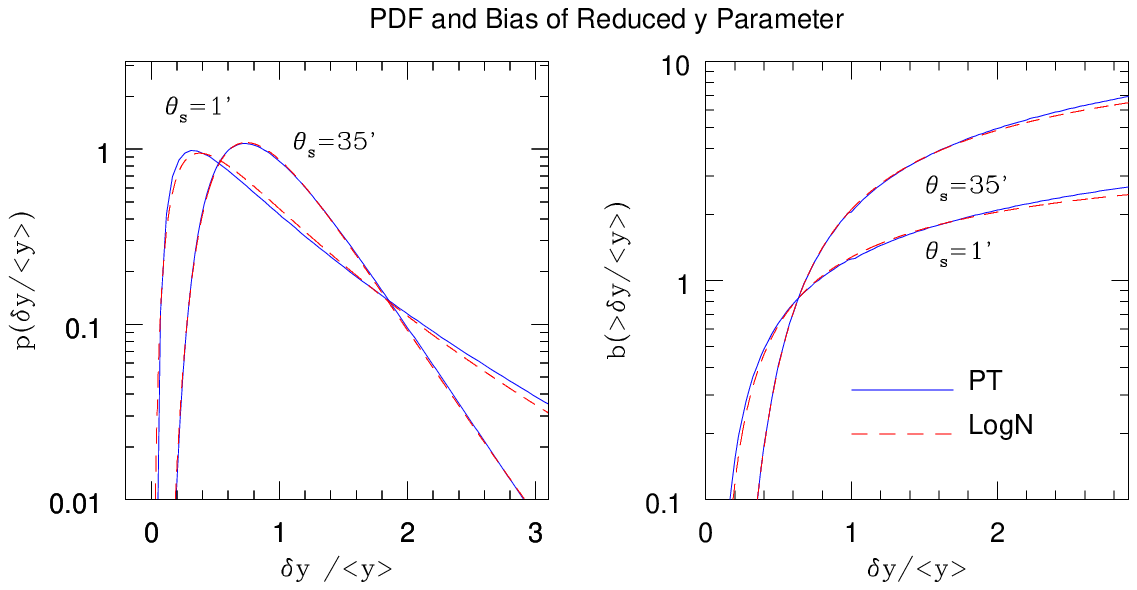}}}
\end{center}
\caption{The PDF defined in \cb{Eq.(\ref{eq:pdf})} (left panel) and bias function defined in 
\cb{Eq.(\ref{eq:cumu_bias})} (right panel) $b(>\delta y/\la y\ra)$ are plotted as a function of $\eta'=\delta y/\la y\ra$.
They are computed using two different approaches. The dashed lines show the
analytical predictions from the hierarchical ansatz and the solid lines show the predictions from
the lognormal distribution. The lines correspond to different angular scales as depicted.}
\label{fig:bias_sz}
\end{figure}
\n
This is a generalization of the usual definition of cumulant correlators for the case of two
different fields in this particular case convergence $\kappa$ and $\delta y$.
The smoothing angular scales and the window function are left completely arbitrary.
These definitions can be also used to define mixed normalised one-point $S^{\kappa y}_N$ parameters (with $N=p+q$)
involving two different fields by considering the limiting situation $\oh_1=\oh_2$.
\begin{eqnarray}
&&\hat{\cal C}_{p+q}\left [[\mI_{\theta_s}]^{p-1}\mI_{\theta_{12}}\right ] = \int_0^{r_0} { \omega_{\rm wl}^p(r) \omega_{\rm sz}^q(r) b_{\pi}^{q}(r)
\over d_A^{2(p+q-1)}(r)}[\mI_{\theta_s}]_{\rm wl}^{(p-1)}[\mI_{\theta_s}]_{\rm sz}^{(q-1)}[\mI_{\theta_{12}}] dr; \quad\quad
[\mI_{\theta_s}]_{\rm wl}  =   \int  \frac{d^2 \bf l}{(2\pi)^2} P_{\delta}
\left( {l \over d_A(r)} \right) W_{\rm TH}^2(l\theta_s);  \\
&& [\mI_{\theta_s}]_{\rm SZ}  =   \int  \frac{d^2 \bf l}{(2\pi)^2} P_{\delta}
\left( {l \over d_A(r)} \right) b_l^2(\theta_s'); \quad\quad
[\mI_{\theta_{12}}]  \equiv  \int
 \frac{d^2\bf l}{(2\pi)^2} P_{\delta} \left( {l \over d_A(r)} \right)
W_{\rm TH}(l \theta_s)b_l(\theta_s') \exp (\; {\cs i}\; \bl \cdot {\bf \theta}_{12}).
\end{eqnarray}
\noindent
Though results can be derived for different smoothing angular scales,
for simplicity we will only consider identical smoothing beam size $\theta_s=\theta_s'$.
The hierarchical expression for the lowest order cumulant
i.e. $S_3^{\kappa}$ for convergence was derived by Hui (1998). He
showed that his result agrees well with numerical ray tracing
experiments. Later studies
have shown that higher order cumulants and even the
two-point statistics such as cumulant correlators
can also be reliably modelled in a similar way (Munshi \& Coles 1999; Munshi \&
Jain 1999a). More recently it was shown \citep{Mu11a} that the statistics of
tSZ too can be modelled according to the same prescription. In particular
it was shown that the lognormal distribution can be used to predict the
PDF and bias associated with the tSZ maps. We extend these
results to the case of joint analysis of weak lensing and tSZ maps.

We will develop these results further to construct the full joint 2PDF
simply using the individual bias functions for tSZ and weak lensing. The hierarchical ansatz
allows us to write the joint 2PDF as:
\be
p^{(i)}_{\kappa y}(\kappa^{(i)},y)d\kappa^{(i)} dy = p^{(i)}_{\kappa}(\kappa^{(i)}) p_{y}(y)( 1
+ b^{(i)}_{\kappa}(\kappa) \xi^{\kappa y}_{12}(\theta_{12}) b_{y}(y)) d\kappa^{(i)} dy,
\ee
\n
and its relation to the bias associated with collapsed objects in
underlying density field $\eta=1+\delta$.

As an aside, it is simple to check that if we are dealing with tomographic redshift slices $\kappa^{(i)}$ and $\kappa^{(j)}$
of the same or different weak lensing surveys the corresponding cumulant correlators are defined as:
\be
\hat{\cal C}_{p+q}^{ij}\left [[\mI_{\theta_s}]_{\rm wl}^{p-1}[\mI_{\theta_s}]_{\rm sz}^{q-1}[\mI_{\theta_{12}}]\right ] = \int_0^{r_0} { \omega_{(i)}^p(r) \omega_{(j)}^q(r) b^q_{\pi}(r)
\over d_A^{2(p+q-1)}(r)}[\mI_{\theta_s}]_{\rm wl}^{(p-1)}[\mI_{\theta_s}]_{\rm sz}^{(q-1)}[\mI_{\theta_{12}}]\; dr.
\ee
\n
The expressions for the cumulant correlators are constructed by replacing $y(\oh)$ in Eq.~(\ref{eq:mixed_cumu})
with $\kappa^{(j)}(\oh)$:
\begin{equation}
 \langle \kappa_{(i)}^p(\oh_1) \kappa_{(j)}^q(\oh_2)\rangle_c  =
 C_{pq}^{\eta}\hat{\cal C}_{p+q}[[\mI_{\theta_s}]^{(p+q-2)}[\mI_{\theta_{12}}]]
= C_{pq}^{\kappa y}\langle\kappa_{(i)}^2\rangle_c^{(p-1)}\langle \kappa_{(j)}^2 \rangle_c^{(q-1)}
\langle \kappa^{(i)}(\oh_1)\kappa^{(j)}(\oh_2)\rangle_c.
\end{equation}
%
%
%

These expressions are quoted here for the highly nonlinear regime.
As is well known a similar hierarchy develops in the quasilinear regime so the same analytical tools are applicable in
such a situation. We will next use these expressions to compute the lower order moment of one and two-point PDFs.

The corresponding results for the Ostriker-Vishniac effect (and kinetic Sunyaev-Zel'dovich (kSZ) effect; \cite{Castro03}) will be presented elsewhere.
\section{Joint Probability Distribution Function $\kappa$ and  $\cal y$}
\label{sec:pdf}
\cs{The primary aim in this section is to prove that with a suitable definition of reduced $y$ parameter, 
its statistics under certain approximation can be reduced to that of the underlying density contrast $\delta$.
The proof depends on a generic hierarchical {\em ansatz}  and a scale independent biasing model. Analogously,
we also define a reduced convergence parameter whose statistics reflects that of underlying mass distribution.
The problem of correlating of $\kappa$ and $y$ then reduces to correlating $\eta$ and $\eta'$ and can be modelled using
techniques developed for analysing correlation structure of the density distribution.}

To compute the bias associated with peaks in the convergence field we
have to first develop an analytic expression for the generating function
$\beta_{\kappa y}(z_1, z_2)$ for the convergence field $\kappa$ and the tSZ field $\delta y=y-\la y\ra$.
For that we will use the usual definition for the mixed two-point
cumulant correlator $C^{\kappa y}_{pq}$:
%
\begin{equation}
C^{\kappa y}_{pq} =
{\langle \kappa^p(\oh_1) \delta y^q(\oh_2) \rangle_c \over  \langle \cs{\kappa}^2 \rangle_c^{p-1} \langle \delta y^2 \rangle_c^{q-1} \langle
\kappa(\oh_1) \delta y(\oh_2) \rangle_c }.
\label{eq:cumu_corr}
\end{equation}
(for a more detailed description of cumulant correlators see Munshi \& Coles, 1999b).
This is a natural generalisation of cumulant correlators found generally in the literature for individual fields \citep{MCV12}.
In the limiting case of $\oh_1=\oh_2$ we recover the limiting case of one point cumulants: $S^{\kappa y}_{p+q} = C^{\kappa y}_{pq}$.
We will show that, like its density field counterpart, the
two-point generating function for the convergence field $\kappa$
can also be
expressed (under certain simplifying assumptions) as a product
of two generating functions $\beta(z)$
which can then be directly related to the bias associated with
``hot-spots'' in the convergence field.
It is clear that the factorization of generating function actually
depends on the factorization property of the cumulant correlators i.e.
$C^{\eta\eta'}_{pq} = C^{\eta}_{p1} C^{\eta'}_{q1}$. Note that such a factorization is
possible when the correlation of two patches in the directions
$\oh_1$ and $\oh_2$ $\two$  is smaller compared to the variance
$\one$ for the smoothed patches:
\be
\beta_{\kappa y}(z_1, z_2) = \sum_{p,q}^{\infty} {C^{\kappa y}_{pq} \over p! q!} z_1^p z_2^q = \sum_{p,q}^{\infty} {1 \over p! q!} { z_1^p
z_2^q\over
\langle \kappa^2 \rangle_c^{p-1} \langle \delta y^2 \rangle_c^{q-1} } {\langle \kappa^p(\oh_1) \delta y^q(\oh_2) \rangle_c  \over \langle
\kappa(\oh_1)\delta y(\oh_2)\rangle_c }.
\ee
Here $z_1$ and $z_2$ are dummy variables. We will now use the integral expression for cumulant correlators
(Munshi \& Coles 1999a) to
express the generating function which in turn uses the hierarchical
{\em ansatz} and the far field approximation as explained above:
\be
\beta_{\kappa y}(z_1, z_2) = \sum_{p,q}^{\infty} {C^{\kappa y}_{pq} \over p! q! } { z_1^p \over
\langle \kappa^2 \rangle_c^{p-1}}{ z_2^q \over \langle \delta y^2 \rangle_c^{q-1} } { 1 \over \xi_{y\kappa}^{12}}  \int_0^{r_0}\; dr\; d_A^2(r)
{\omega_{\rm wl}^{p}(r) \omega_{\kappa}^{q}(r) \over [d_A(r)]^{2p} [d_A(r)]^{2q} }
[\mI_{\theta_s}]_{\rm wl}^{p-1}[\mI_{\theta_s}]_{\rm sz}^{q-1} \mI_{\theta_{12}}.
\ee
It is possible to further simplify the above expression by separating the
summation over dummy variables $z_1$ and $z_2$, which will be useful to
establish the factorization property of two-point generating function
for bias $\beta_{\kappa y}(z_1,z_2)$. We can now decompose the double sum over the two indices into two
separate sums over individual indices.
We do not use any of these specific models for clustering and only
assume the hierarchical nature of the higher order correlation functions. In the past, primarily
data from galaxy surveys have been analysed extensively using these {\em ansatze}. The
motivation here is to show that cross-correlaion statistics of weak-lensing surveys against the
tSZ surveys can also be analysed using such techniques.
The most general result for the lower order cumulant correlators in the large
separation limit can be expressed as:
\be
\beta_{\kappa y}(z_1, z_2) = \inc dr \; d^2_A(r)\; { \mI_{\theta_{12}} \over \xi^{\kappa y}_{12}}
{ \la \kappa^2\ra_c \over {[\mI_{\theta_s}]_{\rm wl}}}
\beta_{\eta} \Big ( { z_1 \over \la \kappa^2 \ra_c} {\omega_{\rm wl}(r) \over d^2_A(r)} [\mI_{\theta_s}]_{\rm wl}  \Big ) { \la \delta y^2 \ra_c
\over {[\mI_{\theta_s}]_{\rm sz}}} \beta_{\eta'} \Big ( { z_2 \over \la \delta y^2 \ra_c} {\omega_{\rm SZ}(r)b_{\pi}(r) \over d_A^2
(r)} [\mI_{\theta_s}]_{\rm sz}  \Big ).
\ee
\begin{figure}
\begin{center}
\begin{minipage}[b]{0.4\linewidth}
{\epsfxsize=6.75 cm \epsfysize=6.5 cm
  {\epsfbox[1 1 339 250]{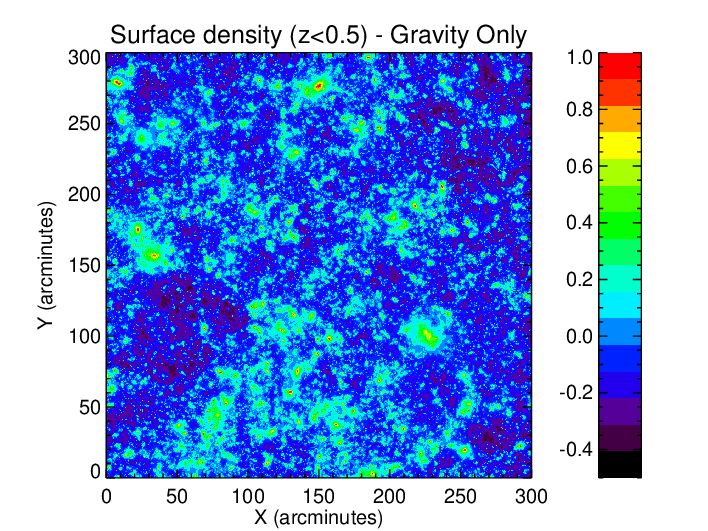}}}
\end{minipage}
\hspace{1.1cm}
\begin{minipage}[b]{0.4\linewidth}
{\epsfxsize=6.75 cm \epsfysize=6.5 cm
 {\epsfbox[1 1 339 250]{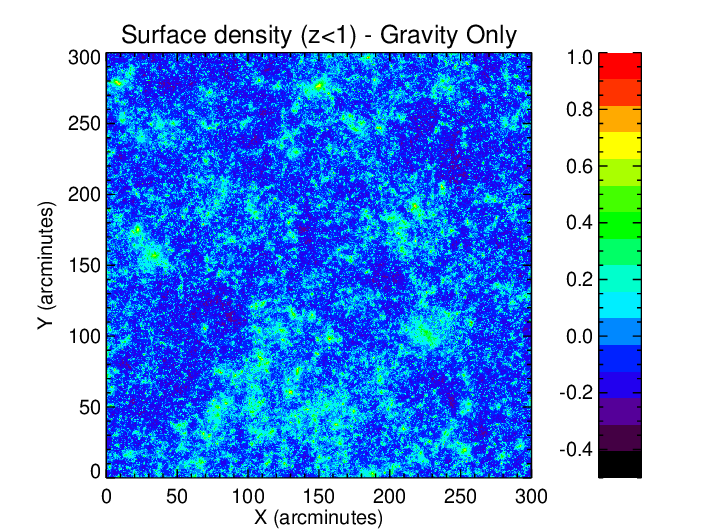}}}
\end{minipage}
\end{center}
\caption{
Simulated $5^{\circ}\times 5^{\circ}$ projected density contrast
maps ${\rm log}_{10}[{1+\delta}]$
are depicted. The maps were generated using Virgo consortium's
Millennium Gas Simulations.
Maps were constructed using line of sight integration out to redshift $z=0.5$ (left panel) and $z=1$ (right panel). We will be using
these projected mass maps as a proxy for weak lensing convergence to test analytical predictions.}
\label{fig:den}
\end{figure}
\n
The above expression is quite general and depends only on the small angle
approximation \citep{Limb54} and the large separation approximation and is valid for any
given specific model for the generating function ${\cal G}_{\delta}(\tau)$. However, it is easy
to notice that the projection effects as encoded in the line of sight
integration do not allow us to write down the two-point generating
function $\beta_{\kappa y}(z_1, z_2)$ simply as a product of two one-point generating functions
$\beta_{\kappa}(z_1)$ and $\beta_y(z_2)$, as generally is the case for the density field $\eta=1+ \delta$.

As in the case of the derivation of the probability distribution
function for the smoothed convergence field $\kappa$, it will be much
easier if we define a reduced smoothed convergence field $\eta_s$. The
statistical properties of $\eta_s$ are very similar to that of the
underlying 3D density field (under certain simplifying approximation) and are roughly
independent of the background geometry and dynamics of the universe. In a similar manner
we also define the reduced $y$ field $\eta_s'$. We define the reduced convergence
$\eta_s$ and the reduced tSZ $y$ map $\eta'_s$ respectively by the following expressions:
$\eta_s = {(\kappa - \kappa_{m})/(-\kappa_{m}}) = (1 + {\kappa
/ |\kappa_{m}|})$ and   $\eta'_s = {\delta y/\la y\ra}$.
Where the minimum \cs{value} of $\kappa_m$ is defined as:
${\cs \kappa}_{m} = -\int_0^{r_s} dr \;\omega_{\rm wl}(r)$ and  $\la y\ra$ is the average value of the $y$ parameter.
It is easy to notice that the minimum value of the convergence field
will occur in those lines of sight which are completely
devoid of any matter, i.e. $\delta = -1 $ all along the line of
sight. We will also find out later that the
cosmological dependence
of the statistics of the $\kappa$ field is encoded in $k_{m}$ and
this choice of the new variable $\eta_s$ makes its related
statistics almost
independent of the background cosmology. Repeating the above analysis
again for the $\eta_s$ field,  we can
express the cumulant correlator generating function for the reduced
convergence field $\eta_s$ as:
\be
\beta_{\eta\eta'}(z_1,z_2) = \inc dr {1 \over |\kappa_m|} {1 \over \la y \ra}  \; d^2_A(r)\; { \mI_{\theta_{12}} \over \xi^{\kappa y}_{12}}
{ \la \kappa^2\ra \over [{\mI_{\theta_s}}]_{\rm wl}}
\beta_{\eta} \Big ( |\kappa_m| { z_1 \over \la \kappa^2 \ra_c} {\omega_{\rm wl}(r) \over d^2_A(r)} [\mI_{\theta_s}]_{\rm wl}  \Big ) { \la \delta y^2 \ra
\over {[\mI_{\theta_s}]_{\rm sz}}} \beta_{\eta'} \Big ( \la y\ra { z_2 \over \la \delta y^2 \ra_c} {\omega_{\rm SZ}(r)b_{\pi}(r) \over d_A^2
(r)} [\mI_{\theta_s}]_{\rm sz}  \Big ).
\ee
While the above expression is indeed very accurate and relates the
generating function of the density field with that of the convergence
field, it is difficult to handle for any realistic practical applications.

\cs{In a scaling analysis the scaling function $h(x)$ can be expressed as an inverse Laplace transform of the
generating function $\phi(z)$ which is a generating function of the normalised one-point cumulants $S_p$ (a deatiled relevant discussion can be found in \cite{Mu11a}): 
\ben
&& h(x) = -{1 \over 2\pi i }\int_{-i\infty}^{i\infty} \exp(x\,z) \phi(z); \quad\quad\quad \phi(z) \equiv \sum_{p} S_p {z^p \over p!}.
\een
Similarly, the bias function $b(x)$ in scaling analysis is related to the function $\tau(z)$. The generating function
defined as $\tau(z)= \sum_p z^p C_{p1}/p!$ acts as a generator for the $C_{p1}$ parameters.
The entire hierarchy of the $C_{pq}$ can be constructed from the $C_{p1}$ parameters i.e. $C_{pq}=C_{p1}C_{q1}$ at large separation limit.
\ben
&& h(x)b(x) = {1 \over 2\pi i }\int_{-i\infty}^{i\infty} \exp(x\,z) \tau(z); \quad\quad\quad  \tau(z) \equiv \sum_{p} C_{p1} {z^p \over p!}.
\een}
It is important to notice that the scaling functions such
as $h(x)$ for the
density probability distribution function and  $b(x)$ for
the bias associated with over-dense objects (with $x = (1+\delta)/\bar\xi^{\delta}_2$) are typically estimated
from numerical simulations especially in the highly non-linear regime ($\bar \xi^{\delta}_2$ is the volume average
of two-point correlation function). Such estimations are plagued by several uncertainties, such as the
finite size of the simulation box. It was noted in earlier studies
that such uncertainties lead to only a rather approximate estimation
of $h(x)$. The estimation of the scaling function associated with the
bias i.e. $b(x)$ is even more complicated due to the fact that the
two-point quantities such as the cumulant correlators and the
bias are more affected by the finite size of the catalogues.
So it is not  fruitful to
actually integrate the exact integral expression we have derived
above and we will replace all line of sight integrals with its
approximate values. The previous study by Munshi \& Jain (1999) have used an exactly similar
approximation to simplify the one-point probability distribution
function for $\kappa$ and found good agreement with ray tracing
simulations. We will show that our approximation reproduces the
numerical results quite accurately for a wide range of smoothing angle,
\begin{eqnarray}
&& |\kappa_{m}| \approx {1\over 2} r_s \omega_{\rm wl}(r_c); \quad\quad
 \la y\ra \approx {1\over 2} r_s \omega_{\rm SZ}(r_c)b_{\pi}(r_c); \quad\quad 0 < r_c < r_s;\\
&&\la\kappa^2\ra_c \approx {1\over 2} r_s  {\omega_{\rm wl}^2(r_c) \over d^2_A(r_c)} \Big [ \int {d^2 {\bf l} \over
(2\pi)^2} { P_{\delta}}({l \over d_A(r_c)}) W^2_{\rm TH}(l \theta_s) \Big ];
\quad\quad \la \delta y^2 \ra_c \approx {1\over 2} r_s  {b^2_{\pi}(r_c)\omega_{\rm SZ}^2(r_c) \over d^2_A(r_c)} \Big [ \int {d^2 {\bf l} \over
(2\pi)^2} { P_{\delta}}({l \over d_A(r_c)}) b^2_{l}(\theta_s) \Big ];\\
&&\two  \approx {1\over 2} r_s  {\omega_{\rm SZ}(r_c)\omega_{\rm wl}(r_c) \over d_A^2(r_c)} \Big [ \int {d^2 {\bf l} \over
(2\pi)^2} { P_{\delta}({l \over d_A(r_c)})} W_{\rm TH}(l\theta_s)b_{l}(\theta_s) \exp [i {\bf l}\cdot \theta_{12}] \Big ].
\end{eqnarray}
Use of these approximations gives us the leading order contributions
to these integrals and we can check that to this order we recover the
factorization property of the generating function i.e. $\beta_{\eta\eta'}(z_1,
z_2) = \beta_{\eta}(z_1) \beta_{\eta'}(z_2)$,
\begin{equation}
\beta_{\eta\eta'}(z_1,z_2) = \beta_{\eta}(z_1) \beta_{\eta'}(z_2) =
\beta_{1+\delta}(z_1) \beta_{1+\delta'}(z_2) \equiv \tau(z_1)\tau(z_2).
\end{equation}
So it is clear that at this
level of approximation, due to the factorization property of the cumulant
correlators, the bias function $b_{\eta}(x)$ associated with the peaks
in the convergence field $\kappa$,  beyond certain threshold, obeys a
similar factorization property too, which is exactly the same as its
density field counterpart. Earlier studies have established such a
correspondence between the convergence field and the density field
in the case of the one-point probability distribution function $p(\delta)$
(Munshi \& Jain 1999b),
\begin{equation}
b_{\eta}(x_1) h_{\eta}(x_1) b_{\eta'} (x_2) h_{\eta'} (x_2) =
 b_{1 + \delta}(x_1) h_{1 + \delta}(x_1) b_{1+\delta'} (x_2) h_{1+\delta'} (x_2).
\end{equation}
The following relation between $\beta_{\eta}(z)$
and $b_{\eta}(x)$ is useful for modelling. However,
for all practical purpose we find that the differential bias $\beta_{>\eta}(y)$
as defined is simpler to measure from numerical
simulations due to its cumulative nature:
\begin{equation}
b_{\eta}(x) h_{\eta}(x) = -{ 1 \over 2 \pi i}
\int_{-i\infty}^{i\infty} dz \tau (z) \exp (xz); \quad
b_{\eta}(>x) h_{\eta}(>x) = -{ 1 \over 2 \pi i}
\int_{-i\infty}^{i\infty} dz {\tau (z)\over z} \exp (xz).
\end{equation}
\begin{figure}
\begin{center}
\begin{minipage}[b]{0.25\linewidth}
\begin{center}
{\epsfxsize=5.5 cm \epsfysize=5.95 cm {\epsfbox[1 1 337 253]{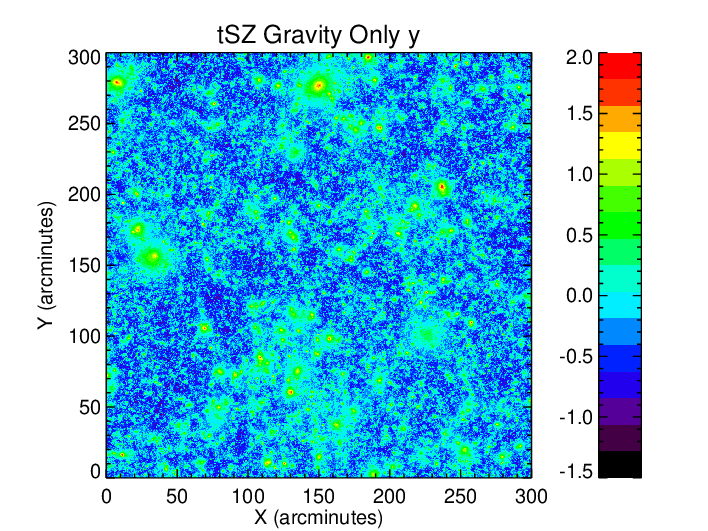}}}
\end{center}
\end{minipage}
\hspace{0.5cm}
\begin{minipage}[b]{0.25\linewidth}
\begin{center}
{\epsfxsize=5.5 cm \epsfysize=5.95 cm {\epsfbox[1 1 337 253]{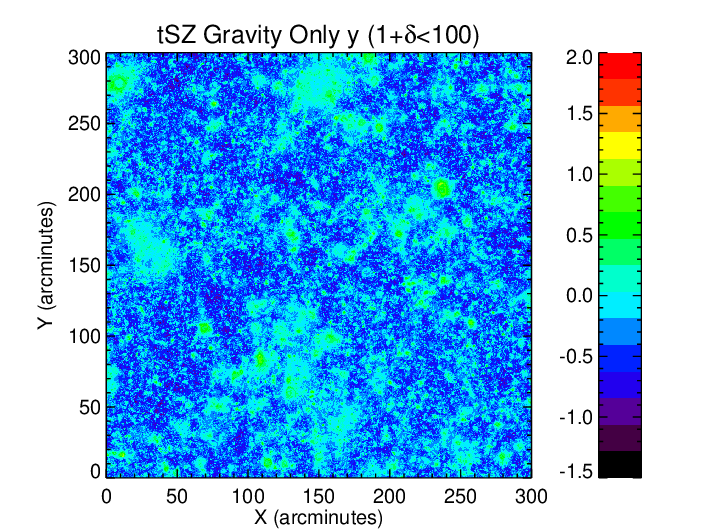}}}
\end{center}
\end{minipage}
\hspace{0.5cm}
\begin{minipage}[b]{0.25\linewidth}
\begin{center}
{\epsfxsize=5.5 cm \epsfysize=5.95 cm {\epsfbox[1 1 337 253]{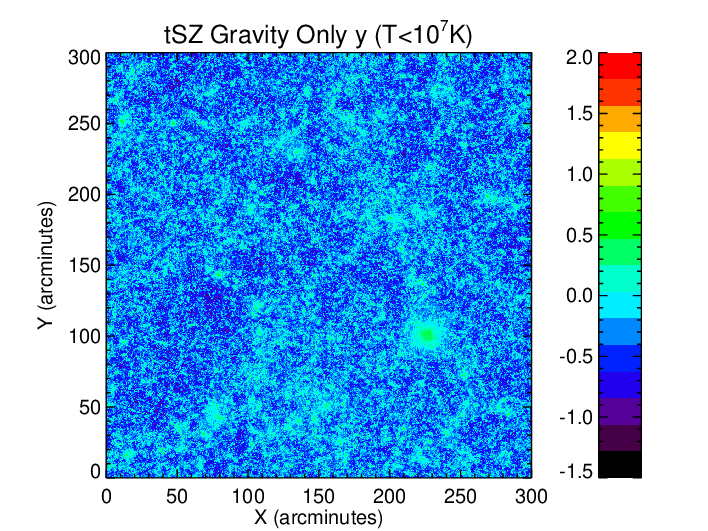}}}
\end{center}
\end{minipage}
\end{center}
\caption{Simulated $5^{\circ}\times 5^{\circ}$ dimensionless
scaled thermal Sunyaev-Zel'dovich maps ${\rm log}_{10}[{y/\la y\ra}]$
are depicted. The maps were generated using Virgo consortium's 
Millennium Gas Simulation. The left panel shows the resulting $y$ map.
The middle panel correspond to
map generated using low density regions and will be referred to as a $y_{\rho}$ map. Only over dense regions
with density $1+\delta <100$ were considered.
The right panel correspond to low temperature regions ${\rm T} < 10^7{\rm K}$ and will be referred to as a $y_{\rm T}$ map.
These set of hydrodynamic simulations ignore pre-heating but takes into account
adiabatic cooling. We will refer to them as GO or Gravity-Only simulations (see text for
more details). We use $\la y \ra$ of the total signal when
showing the results for low density and temperatures. }
\label{fig:tsz_nc}
\end{figure}
\begin{figure}
\begin{center}
\begin{minipage}[b]{0.25\linewidth}
\begin{center}
{\epsfxsize=5.5 cm \epsfysize=5.95 cm {\epsfbox[1 1 337 253]{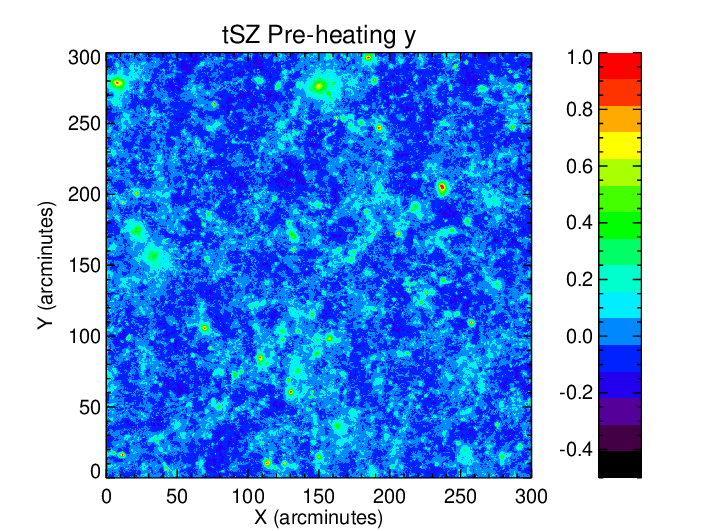}}}
\end{center}
\end{minipage}
\hspace{0.5cm}
\begin{minipage}[b]{0.25\linewidth}
\begin{center}
{\epsfxsize=5.5 cm \epsfysize=5.95 cm {\epsfbox[1 1 337 253]{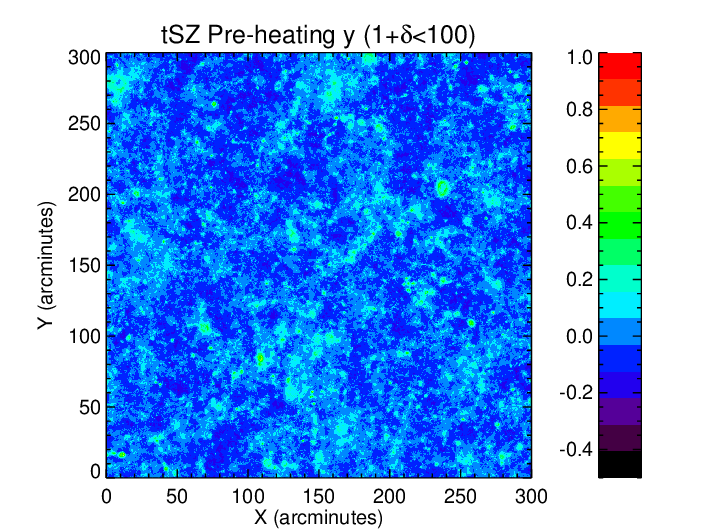}}}
\end{center}
\end{minipage}
\begin{minipage}[b]{0.25\linewidth}
\begin{center}
{\epsfxsize=5.65 cm \epsfysize=6.15 cm {\epsfbox[-30 -10 337 253]{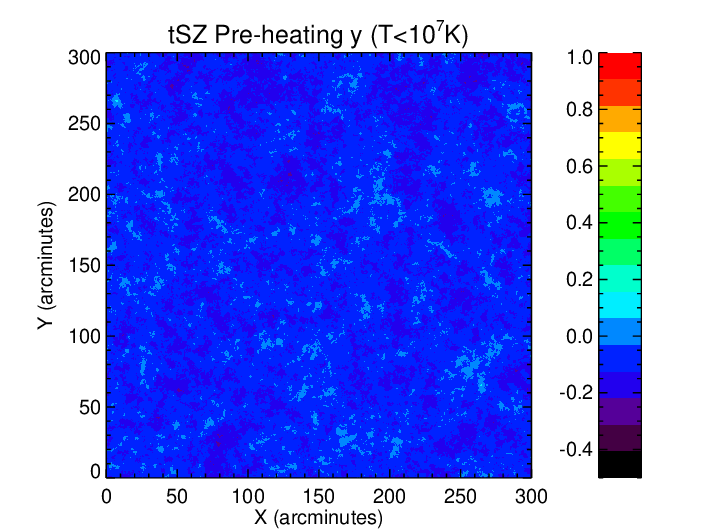}}}
\end{center}
\end{minipage}
\end{center}
\caption{Same as previous figure, but for simulations with pre-heating and cooling.
These simulations will be referred to as PC simulations.}
\label{fig:tsz_pc}
\end{figure}
%
Using these results we can finally write down:
\be
p_{\eta\eta'}(\eta,\eta')d\eta d\eta' = p_{\eta}(\eta)p_{\eta'}(\eta')(1+b_{\eta}(\eta)\xi^{\eta\eta'}_{12}b_{\eta'}(\eta'))d\eta d\eta';
\quad \eta = \cs{\kappa_s/|\kappa_m|}; \quad \eta' = \delta y/\la y\ra; \quad\quad \xi^{\eta\eta'}_{12}=\la\eta(\oh_1) \eta'(\oh_2)\ra.
\label{eq:2point}
\ee
It is important to notice that although the bias $b_{\eta}(x)$
associated with the reduced convergence field  and the underlying density
field are exactly same, the variance associated with the density field is
very high but the projection effects in the convergence field
bring down the variance in the convergence field to less than unity
which indicates that we have to use
the integral definition of bias to recover it from its generating
function (see Appendix A for more details).
Finally, writing down the joint probability distribution function $p_{\kappa y}(\kappa, y)$
for the smoothed projected convergence field $\kappa$ and the $y$
in terms of their reduced versions $\eta$ and $\eta'$:
\ben
&& p_{\kappa y}(\kappa, y)d\kappa dy = p_{\kappa}(\kappa) p_{y}(y)( 1
+ b_{\kappa}(\kappa)\;\xi^{\kappa y}_{12}\; b_{y}(y)) d\kappa\;dy \nn ; \label{eq:2pdf}\\
&& b^{}_{y}(y) = b_{\eta}(y/\la y \ra)/ {\la y \ra}; \quad
b^{}_{\kappa}(\kappa) = {b_{\eta}(\kappa/|\kappa_m|)/ {|\kappa_{m}|}};\quad\quad
\xi^{\kappa y}_{12} \equiv \la\kappa(\oh_1) y(\oh_2) \ra ={\xi^{\eta\eta'}_{12}/ (|\kappa_{m}| \la y\ra}).
\label{eq:joint_pdf}
\een
In an earlier study, Munshi et al. (1999b) used similar arguments for the convergence maps
to show that $p_{\kappa}(\kappa) = {p_{\eta}(\kappa/|\kappa_m|)/ |{\cs{\kappa}_{m}}|}$ and recently it was
generalised by (Munshi et al. 2011) for the $y$-maps to $p_{y}(y) = {p_{\eta}(y/\la y\ra)/ {\la y\ra}}$
If we notice that $\xi^{\kappa y}_{12} =
{\xi^{\eta\eta'}_{12}/ (|\kappa_{m}|\la y \ra)}$; then the above expressions helps us to
write $b_{\kappa}(\kappa) = {b_{\eta}(\kappa/|\kappa_m|)/ {|\kappa_{m}|}}$ and $b_{y}(y) = {b_{\eta}(y/\la y\ra)/ {\la y\ra}}$.

\cs{Eq.(\ref{eq:joint_pdf}) is the main result of our analysis - it provides an expression for the joint PDF
of $\kappa$ and $y$ which encapsulates knowledge of cumulant correlators of all orders. 
We have shown that we can define two variables $\eta$ (reduced convergence) and $\eta'$ (reduced $y$ parameter)
that can simplify the derivation considerably. Assuming a scale independent biasing model we can
directly link the statistics of $y$ parameter to that of the underlying density contrast $\delta$.
Using these variables, we have shown that the PDF and bias of $y$ can be directly linked to that of $\delta$.  
The mixed lower order cumulant correlators involving $\kappa$ and $y$ maps can then be expressed interms of the 
bias functions of $\kappa$ and $y$. We also stress here that presence of significant non-gravitational effect 
will violate the assuptions that we have used in our analytical derivation.}

A similar result can of course be derived for cross-correlation of tSZ $y$ map and weak lensing
convergence maps $\kappa^{(i)}$ from individual tomographic bins. The joint PDF $p^{(i)}(\kappa^{(i)},y^{})$
for tomographic maps $\kappa^{(i)}$ and projected $y$ maps can be expressed in terms
of the individual PDF $p^{(i)}(\kappa)$, $p^{}(y)$ maps and the bias $b^{(i)}_{\kappa}(\kappa)$
and $b^{}_{y}(y)$:
\begin{eqnarray}
&& p^{(i)}(\kappa^{(i)},y^{})d\kappa^{(i)} \; dy = p^{(i)}(\kappa) p^{}(y)( 1
+ b^{(i)}_{\kappa}(\kappa^{(i)})\; \xi^{(i)}_{12}\; b^{}_{y}(y)) d\kappa^{(i)} dy; \nn \\
&& b^{(i)}_{\kappa}(\kappa) = {b_{\eta}(\eta)/ {|{\cs \kappa}^{(i)}_{m}|}}; \quad
b_{y}(y) = b_{\eta}(\eta')/ {\la y \ra}; \quad \xi_{12}^{(i)} \equiv
{\xi^{\eta\eta'}_{12}/ (|\kappa_m^{(i)}|\la y\ra)}.
\label{eq:joint_tomography}
\end{eqnarray}
Notice that the bias and one point PDF for both the projected convergence maps $\kappa$
and tomographic maps $\kappa^{(i)}$ as well as for the tSZ map $y$ are all constructed
from the same underlying PDF $p_{\eta}$ and the bias $b_{\eta}$ associated with the
3D density distribution. The individual PDF $p_{\kappa}(\kappa)$ and $p_y(y)$ and the bias
$b_{\kappa}(\kappa)$ and $b_y(y)$ have already been tested against numerical simulations
and were found to be remarkably successful.

This technique will also allow for the computing of joint PDF of convergence maps for CMB lensing
and the tSZ $y$ maps. This can be achieved by replacing the $\kappa_m$ in Eq.~(\ref{eq:joint_pdf}) with
corresponding $\kappa_m$ for the last scattering surface (LSS), i.e. $\kappa_m = -\int_0^{r_{\rm LSS}}dr w_{\rm wl}(r)$ where
the source distance $r_{s}$ is now replaced by the comoving distance to the LSS $r_{\rm LSS}$.
This is especially relevant given the recent evidence for a CMB lensing signal with WMAP and ACT \citep{Das11, Smidt11}
The results for the cumulant correlators can be modified analogously.
Though the results Eq.~(\ref{eq:joint_pdf}) and Eq.~(\ref{eq:joint_tomography}) are derived using
the hierarchical ansatz the final results are remarkably independent of details of any of the
assumptions that were used in deriving them. This is an indication of their more generic
validity. Indeed it has been shown that the PDF and the bias for the tSZ and weak lensing
fields can be constructed from equivalent but other valid descriptions of PDF and bias
of underlying 3D density contrast. The lognormal distribution for the underlying PDF and
bias provide one such description. In Appendix-B we have provided a short description of
the lognormal distribution. The lognormal model based PDF and bias for $\kappa$ and $y$ are given by:
\be
p_{\kappa}(\kappa) = {p_{\ln}(\kappa/|\kappa_m|)/ {|\cs{\kappa}^{}_{m}}|}; \quad
p_{y}(\delta y) = p_{\ln}(\delta y/\la y\ra)/ {\la y \ra}; \quad
b_{\kappa}(\kappa) = {b_{\ln}(\kappa/|\kappa_m|)/ {|\cs{\kappa}^{}_{m}|}}; \quad
b_{y}(\delta y) = b_{\ln}(\delta y/\la y \ra)/ {\la y \ra}.
\label{eq:beta}
\ee
The resulting PDF and bias matches with the ones from perturbative calculations
for higher smoothing angular scales.

As mentioned earlier, the joint probability of $p_{\kappa y}(\kappa,y)$ is a noisy statistics. The
integrated measure or cumulative PDF of the convergence crossing a threshold $\kappa$ and the tSZ field $y$
crossing a threshold  $\kappa_0$ and $y_0$ respectively is given by:
\ben
&& p_{\kappa y}(>\kappa_0,>y_0) = \int_{\kappa_0}^{\infty} d\kappa \int_{y_0}^{\infty} dy \; p_{\kappa y}(\kappa,y);\quad
 p_{\kappa}(>\kappa_0)=\int_{\kappa_0}^{\infty} d\kappa p_{\kappa}(\kappa); \quad p_y(>y_0)=\int_{y_0}^{\infty} dy p(y);
\label{eq:pdf}\\
&& b_{\kappa}(>\kappa_0)=\int_{\kappa_0}^{\infty} d\kappa p(\kappa) b_{\kappa}(\kappa) \Big /
\int_{\kappa_0}^{\infty} d\kappa p(\kappa); \quad\quad
b_y(>y_0)=\int_{\kappa_0}^{\infty} d\kappa p_{y}(y) b_{y}(y) \Big / \int_{y_0}^{\infty} dy p_{y}(y).
\label{eq:cumu_bias}
\een

The following statistics can be constructed from the cumulative bias  to probe the joint SZ and weak-lensing cross-correlation
to all orders:
\be
{\cal B}(>\kappa_0,>y_0) = [b_{\kappa}(>\kappa_0)b_y(>y_0)]^{1/2}= {1 \over \sqrt{\xi_{12}^{\kappa y}}}
\left [ {p_{\kappa y}(>\kappa_0,>y_0) \over p_{\kappa}(>\kappa_0)p_y(>y_0)} -1 \right ]^{1/2}
\label{eq:def_b}
\ee
It is important to notice that our modelling of the PDF $p_\kappa$ and $p_y$ is independent of the
modelling of respective variances. The variance calculations for $y$ depend on
inputs such as the detailed modelling of bias and its redshift evolution that
can be independently checked by comparing with simulations.
However, given a correct variance the lognormal model or the hierarchical ansatz
can be used to model the entire PDF as we have shown.
For construction of the PDF of the scaled or reduced
tSZ i.e. $\eta$ we also need the average $\la y \ra$. The modelling of average
$\la y \ra$ is independent of the construction of PDF. However what we have
shown here is that given these two inputs we can reliably predict the
PDF of $y$. While the PDF of reduced tSZ field $\eta$ is independent of
cosmology the dependence of $y$ parameter is encoded in the definition
of reduced $y$ field $\eta' = \delta y/\la y\ra$. The convergence PDF
is defined in terms of the PDF for the reduced convergence $\eta = \kappa/|\kappa_m|$. The PDF
for the reduced convergence $\kappa$ as well as the scaled tSZ map $y$ are both
shown to be same as the PDF of underlying ark matter distribution $\delta$.
The dependence of $p_\kappa$ on cosmological parameters is encoded
in the definition of $\kappa_m$. The cosmological
parameter dependence of bias $b_{\kappa}(\kappa)$ and $b_{y}(y)$ is encoded in
$\la y\ra$ and $\kappa_m$.

It is also interesting that our calculations show that the PDF of {\it projected} or 2D convergence
as well as the tSZ maps are described by the underlying 3D density contrast $\delta$ which
is in agreement with \citep{MuJa00,MuJai01} and provides robust mathematical justification
for the use of lognormal model and its link to linear biasing prescription.

A simple order of magnitude estimate for the weak lensing cumulants is $S_p^{\kappa} = \la \kappa^p_s\ra/\la \kappa^2_s \ra^{p-1}$
is given by $S_p^{\kappa} = S_p/| \kappa_m |^{p-2}$. The cumulants of the field $\delta y/\la y\ra$ simply follows
that of the underlying density contrast. The normalised cumulant correlators of the smoothed convergence field $\kappa$
and the field $\delta y/\la y\ra$ is given by $C_{pq}^{\kappa y}= C_{pq}^{\eta\eta'}/|\kappa_m|^{p-1}$.
\begin{figure}
\begin{center}
{\epsfxsize=15.5 cm \epsfysize=5. cm {\epsfbox[21 530 587 712]{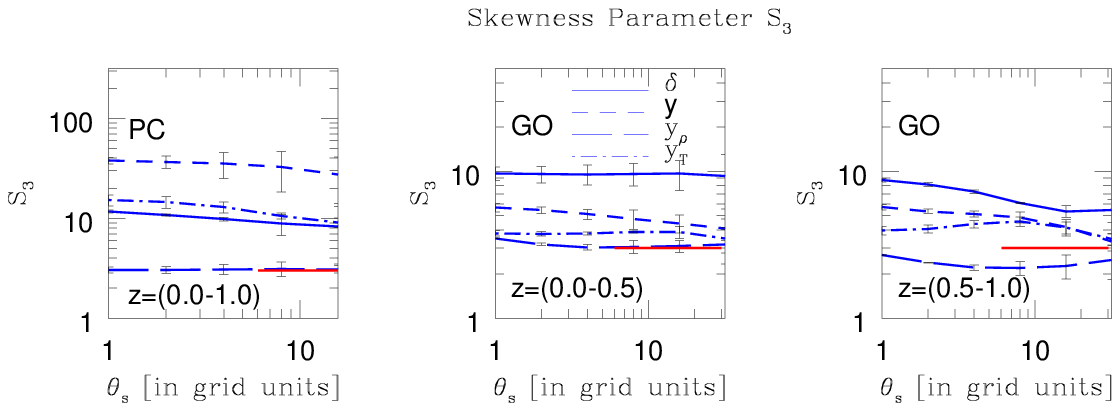}}}
\end{center}
\caption{The skewness parameter $S_3 = \la\delta^3\ra_c /\la\delta^2\ra^2$, signifying the lowest-order departure from Gaussianity is plotted as a function of 
the smoothing angular scale $\theta_s$ for various simulations. The size of the grid interval is 15''.
The left-panel shows the results from PC simulations.
The middle and right panel correspond to GO simulations. The redshift range considered is $z=(0.0-\cs{1.0})$ for the PC simulations.
Two different redshift range are considered for the GO simulations.
The middle-panel correspond to $z=(0.0-0.5)$ and the right-panel correspond to $z=(\cs{0.5}-1.0)$.
The scatter or error-bars are computed using three realisations for each 
simulation type. Various lines correspond to density contrast $\delta$, $y$ parameter (short-dashed lines), 
$y$ parameter contributed only by
the low-temperature phase (long-dashed lines) and the low-density phase (dot-dashed lines) respectively.
The smooth solid lines in middle and right panels correspond to the log-normal prediction 
for $S_3$ i.e. $S_3 = 3$. The finite survey size not only introduces scatter or variance but it also
introduces a bias in the estimator (See text for more details.)}
\label{fig:skew}
\end{figure}

\begin{figure}
\begin{center}
{\epsfxsize=15.5 cm \epsfysize=5. cm {\epsfbox[21 530 587 712]{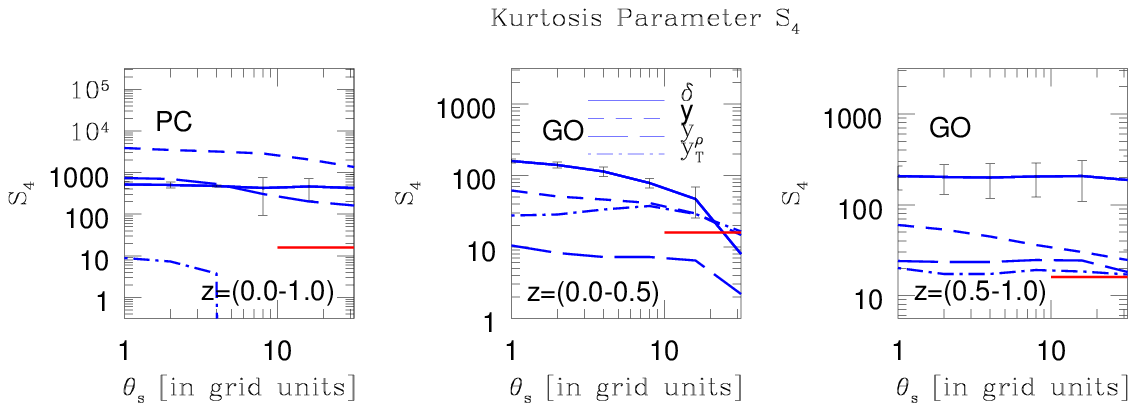}}}
\end{center}
\caption{The kurtosis parameter $S_4 = \la \delta^4\ra_c/\la\delta^2\ra^3$ and similarly for other maps, signifying the next to leading-order departure from Gaussianity is plotted as a function of 
the smoothing angular scale $\theta_s$ for various simulations (in grid units). The size of the grid is 15''.
The left-panel shows the results from PC simulations.
The middle and right panel correspond to GO simulations. The redshift range considered is $z=(0.0-\cs{1.0})$ for the PC simulations.
Two different redshift ranges are considered for the GO simulations.
The middle-panel corresponds to the redshift $z=(0.0-0.5)$ and the right-panel corresponds to $z=(\cs{0.5}-1.0)$.
The scatter (or error-bars) are computed using three realisations for each 
simulation type. Various lines correspond to different maps we have studied i.e. density contrast $\delta$ (solid lines), $y$ parameter (short-dashed lines), $y$ parameter contributed only by
the low-temperature phase (long-dashed lines) and the low-density phase (dot-dashed lines) respectively.
The smooth solid lines in middle and right panels correspond to the log-normal prediction 
for $S_4$ at large smoothing angular scale i.e. $S_4 = 16$ (See text for more details). }
\label{fig:kurt}
\end{figure}
\begin{figure}
\begin{center}
{\epsfxsize=15.5 cm \epsfysize=5. cm {\epsfbox[21 530 587 712]{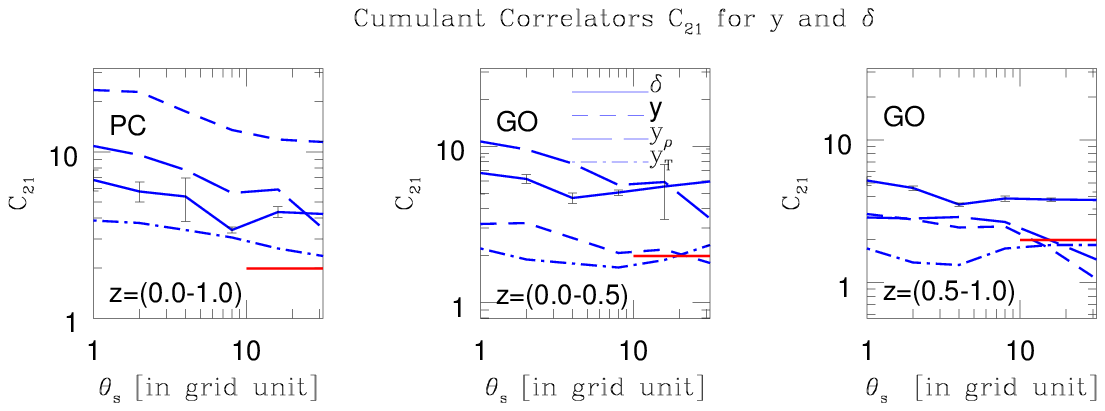}}}
\end{center}
\caption{The normalised cumulant correlators $C^{\delta\delta}_{21}\equiv \la\delta_1^2\delta_2\ra_c/\la\delta_1^2\ra\la\delta_1\delta_2\ra $ 
defined in Eq.(\ref{eq:cumu_corr}) where ($\delta_1\equiv \delta(\oh_1)$) and equivalently for other maps, are plotted as a function of smoothing angular scale $\theta_0$. The left panel
shows results from PC simulations. The other two panels correspond to GO simulations. The solid line corresponds to $C^{\delta\delta}_{21}$. The small dashed lines
correspond to $C^{yy}_{21}$. The long-dashed lines and dot-dashed lines are results computed from low-density ($\rho$)
and low-temperature \,($\rm T$) maps respectively. The middle and right panels correspond to GO simulation with tomographic bins
$z=(0.0-0.5)$ and $z=(0.5-1.0)$ respectively. The errors-bars are computed using three realisations. The angular separation $\theta_{12}$ is fixed at
$\theta_{12}=4\theta_s$. The cumulant correlators are two-point statistics and are more susceptible to finite volume effects.
The solid line at lower right hand corner represent the analytical predictions from lognormal approximation $C_{21}=2.0$.}
\label{fig:cumu1}
\end{figure}
\begin{figure}
\begin{center}
{\epsfxsize=15.5 cm \epsfysize=5.2 cm {\epsfbox[21 530 587 742]{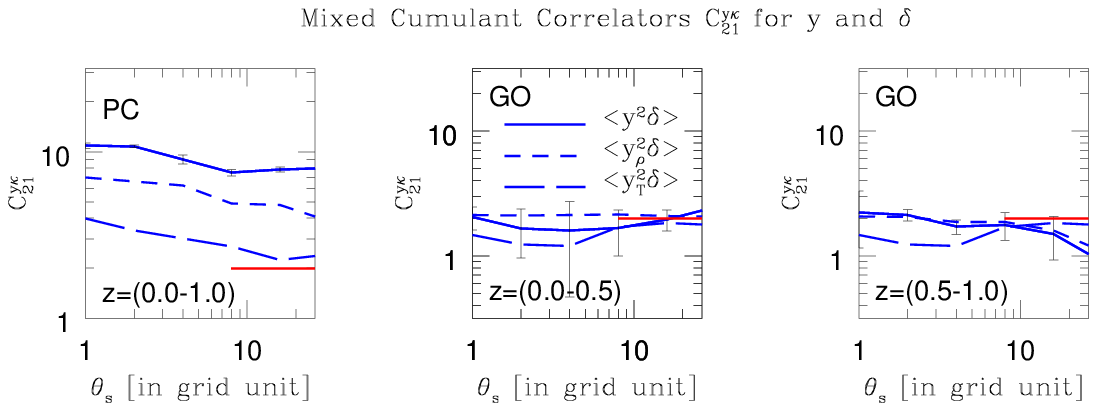}}}
\end{center}
\caption{The normalised {\em mixed} cumulant correlators $C^{y\kappa}_{21} = \la y^2\kappa\ra_c /\la y^2 \ra \la y_1\kappa_2\ra$
(defined in Eq.(\ref{eq:cumu_corr})  are plotted as a function of smoothing angular scale $\theta_0$. The left panel
shows results from PC simulations. The other two panels correspond to GO simulations at different redshift-bins. The solid line corresponds to $C^{y\kappa}_{21}$. The short- and long-dashed lines are results computed from $y_\rho$
and $y_{\rm T}$ maps respectively. The error-bars are computed using three realisations. 
The angular separation $\theta_{12}$ is fixed at
$\theta_{12}=4\theta_0$. The cumulant correlators are two-point statistics and are more susceptible to finite volume effects.
The solid line at lower right hand corner represent the analytical predictions from lognormal approximation $C_{21}=2.0$.
Notice that we use the projected density $\delta$ as a proxy for weak lensing convergence $\kappa$.}
\label{fig:cumu2}
\end{figure}
\begin{figure}
\begin{center}
{\epsfxsize=15.5 cm \epsfysize=5.2 cm {\epsfbox[21 530 587 742]{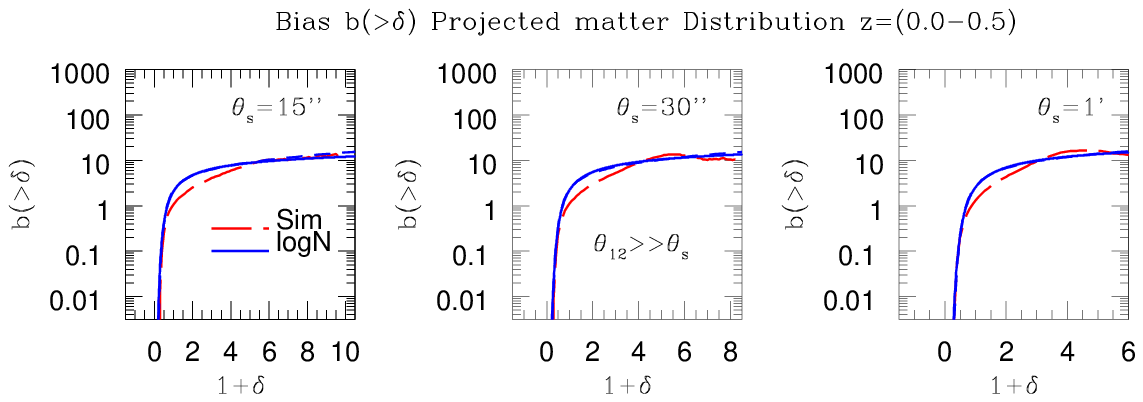}}}  
\end{center}
\caption{The cumulative bias $b(>\delta)$ defined in Eq.(\ref{eq:2point}) as a function of $\delta$. Three different smoothing
angular scales are considered $\theta_s=\cs{15}''$ (left-panel), $\theta_s=\cs{30}''$ (middle-panel) and $\theta_s=1'$ (right-panel) 
respectively. The results from the simulations are plotted in long-dashed lines. The log-normal results
are shown using solid lines. Different levels of dilution of the data were considered.}
\label{fig:bias1}
\end{figure}

\cb{The numerical results for the cumulative bias for reduced convergence $b(>\eta)$ are shown in Figure \ref{fig:bias_kappa}. In the left panel we show 
results for smoothing angular scale $\theta_s=10'$ and in the right panel we show results for  $\theta_s=5'$ The
results are displayed for different redshift bins. Lower curves corresponds to lower redshift bins. We have presented
results for both lognormal distribution as well as perturbative calculations.}

\cb{The corresponding results for the reduced y-parameter is shown in Figure \ref{fig:bias_sz}. The left panel shows the PDF for $\delta y$
and the right panel shows the cumulative bias. The angular scales are $\theta_s=1'$ and $\theta_s=35'$. Both results of perturbative 
calculations and lognormal approximations are shown in each panel.}

\section{Hydrodynamical (SPH) Simulations}
\label{sec:hydro}
The simulated y-maps that we have used were generated by \cite{Scott12}
using {\em millennium gas simulations}  \citep{Hart08,SRE09,Young11, Short10}. Which in turn were
generated to provide hydrodynamic versions of the Virgo consortium's
\footnote{http://www.virgo.dur.ac.uk/}
\footnote{http://www.mpa-garching.mpg.de/galform/millennium/}
dark matter Millennium Simulations and were performed using
publicly-available GADGET2 N-body/hydrodynamics code
\citep{Spring05}. Two different versions of the simulations use same
initial conditions and box-size. In the first run, the gas was
modelled as ideal non-radiative fluid and was allowed to go
adiabatic changes in regions of non-zero pressure gradient. The
evolution was modelled using smooth particle hydrodynamics (sph). An
artificial viscosity too was used to convert bulk kinetic energy of
the gas into its internal energy. This is essential to capture the
physics of shock and thus generate quasi-hydrostatic equilibrium.
These process ensures quasi-hydrostatic equilibrium inside
vitalized halos. See text for more details of the hydrodynamic
simulations used to generate these maps. These set of simulations
will be referred as Gravity Only (GO) simulations. Non radiative
descriptions of inter-cluster gas do not reproduce the observed X-ray
properties of the clusters \citep{Voit05}. So the next set of
simulations that we use pre-heated gas at high redshift that can
generate the required core entropy and capable of producing a
steeper X-ray luminosity-temperature in agreement with observations
. The entropy level of these second set of simulations were chosen
to match the mean X-ray luminosity temperature relation at $z=0$
\citep{Kaiser91,EH91}. These simulations also include radiative
cooling and an entropy sink. We will refer to these simulations as
PC. Cooling in these simulations do not play an important role as
the cooling time for the preheated gas is long compared to the
Hubble time. The cosmological parameters of these simulations are
$\Omega_{\rm CDM}=0.25$, $\Omega_{\Lambda}=0.75$, $\Omega_b =0.045$,
$h=0.73$ and $\sigma_8=0.9$.

\begin{figure}
\begin{center}
{\epsfxsize=15.5 cm \epsfysize=5. cm {\epsfbox[21 525 587 712]{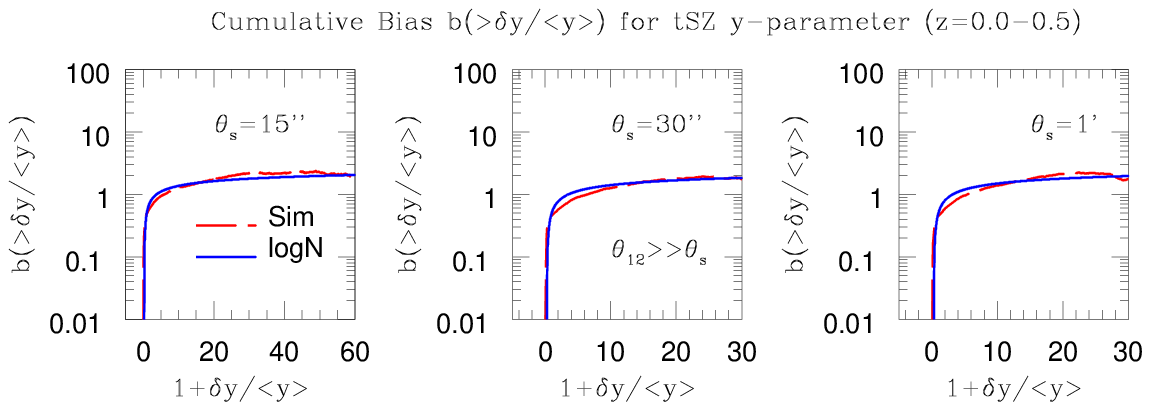}}}
\end{center}
\caption{The cumulative bias $b(>\delta y/\la y\ra)$ defined in Eq.(\ref{eq:joint_tomography}) is plotted as a function of 
$\delta y/\la y \ra$ for the tSZ $y$-parameter (GO simulations). Three different smoothing
angular scales are considered $\theta_s=\cs{15}''$ (left-panel), $\theta_s=\cs{30}''$ (middle-panel) and $\theta_s=\cs{1}'$ (right-panel) 
respectively. The simulation results are plotted with long-dashed lines. The log-normal results
are shown with solid lines respectively. Different levels of dilution of the data were considered.
The variance associated with y-maps is higher compared to the $\delta$ maps which explains 
larger dynamic range for $\delta y$.}
\label{fig:bias2}
\end{figure}
The projected mass distribution for 
both GO and PC simulations are presented in  Figure \ref{fig:den}.
The scaled  $\log_{10}[{y/\la y \ra}]$ parameter distribution of a
realisation is shown in Figure \ref{fig:tsz_nc} for gravity only or GO
simulations and Figure \ref{fig:tsz_pc} for simulations with
pre-heating and cooling (PC). 
The left panels show contribution from
all individual components. The middle panels represents contribution
from over dense regions that satisfy the constraint $1+\delta <
100$. Finally the right panels correspond to the contribution to the
y-map from gas which satisfy the constraint ${\rm T}<10^5{\rm K}$.

There is a very clear and obvious difference between the two sets of
maps in that the GO maps have more substructure. The smoothness of
the PC maps is due to the external thermal energy added to the gas
by the pre-heating process. The mean y-parameter in the GO
simulations is $\la y\ra=2.3\times10^{-6}$ and in the PC simulations
it is nearly four times higher $\la y\ra=9.9\times10^{-6}$. These
values are consistent with COBE/FIRAS constraint $\la y \ra \le 1.5
\times 10^{-5}$. However, it is believed such a high level of
preheating would definitely remove some of the absorption features
seen in the Lyman-$\alpha$ spectrum observed towards quasars \citep{TMS01,SCH07,BV09}. Indeed
the PC model studied here should be treated as an {\em extreme test}
of the effect of a high pre-heating scenario.

In terms of source contributions, the bulk of the $y$-signal comes
from low redshift i.e. $z<2$. However in case of PC simulations the
opposite is true, where $80\%$ of the signal originates from
$z<3.5$. The overdense regions such as the group or clusters are the
sources of $y$-signal in the GO simulations which are primarily
embedded in structures that collapsed at relatively lower redshift.
In case of the PC simulations most of the signals comes from mildly
overdense gas at high redshift. It's interesting to notice that the
GO simulations do get contributions from the gas at high redshift
$z>4$. However such contributions are completely erased in case of
the PC simulations. This is primarily due to the fact that radiative
cooling erases most of the ionized gas at these redshifts.
Due to unavailability of joint tSZ and ray-tracing simulations for weak-lensing,
we have used the projected density as a proxy for weak-lensing convergence.
Indeed weak lensing convergence is identical to projected density contrast
with a redshift dependent weight factor \citep{MuJa00,MuJai01}. 

\section{Tests against Numerical Simulation}
\label{sec:disc}
In this section we provide a detailed discussion of the statistics we employed to analyse the simulation.

{\bf Cumulants and Cumulant Correlators:} We have computed the lower order moments (cumulants)
namely the skewness parameter $S_3$ and the kurtosis parameter $S_4$
from both the GO and PC simulations. These results are presented 
in Figure \ref{fig:skew} (skewness) and  Figure \ref{fig:kurt} (kurtosis)
respectively. The range of smoothing angular scale $\theta_s$ probed corresponds roughly to $15''-2'$.
We find that at angular scales $\theta_s>1'$ the skewness computed 
from GO simulation matches very closely with analytical predictions.
In addition to the skewness parameter the kurtosis parameter
too matches with theoretical predictions at smoothing angular scales
 $\theta_s>1'$.  For both redshift bins we have tested the numerical results
are close to lognormal predictions $S_3=3$ and $S_4=16$. While the higher order moments are
reproduced well in GO simulations, the results from PC simulations
where non-Gravitational effects are dominant, do not match with
theoretical predictions. This is consistent with our previous studies
where we found that only in case of the GO simulations,
where gravitational dynamics and adiabatic cooling are the main
factors influencing structure formation ,the numerical PDF is reasonably
reproduced by theoretical predictions. The lognormal prediction
starts to break down at smaller scale and the linear biasing scheme too
is less accurate at scales smaller than few arc-minutes. 
We have used three realisations to compute the scatter in
$S_3$ and $S_4$. Notice that small survey size can introduce not only a scatter but a significant
bias in estimation of $S_{\rm N}$ parameters in general at large $\theta_s$. We have used a $1024\times 1024$
grid to compute the higher order moments. The
lowest probability that we can resolve using this 
grid is roughly $10^{-6}$. For larger $\theta_s$, the number is
higher. 

In addition to the cumulants for the $y$ parameter, we have also studied the 
cumulant correlators $C^{yy}_{21}$ defined in Eq.(\ref{eq:cumu_corr}) for $y$ parameter, and presented
in Figure \ref{fig:cumu1}. The mixed cumulant correlator $C^{y\kappa}$ is presented in Figure \ref{fig:cumu2}.
The results for cumulants and their correlators follow a very similar pattern. The relevant
scales for cumulant correlators are the smoothing angular scale $\theta_0$
and the separation angular scale $\theta_{12}$. The perturbative results are typically obtained 
in the limiting case of $\theta_{12}\gg \theta_s$. For numerical
implementation we have chosen $\theta_{12} \approx 4\theta_0$.
It is important to realise that being two-point statistics the
cumulant correlators are affected a lot more by the finite volume
effect than their one-point counterparts. The theoretical predictions
for the lowest order predictions for cumulant correlator for lognormal distribution is $C_{21}=2$.
The theoretical results match reasonably well with numerical results
for GO simulations. Both cumulants and cumulant correlators can be
used to differentiate the effect of non-gravitational processes as well
as for sanity checks.

{\bf Contributions from Individual Components:}\;In addition to studying the $\hat y$ maps, we have divided the
entire contribution from various baryonic components, to check how
our theoretical prescriptions compare with that from simulations for
individual components. In this context, we notice that,
thermodynamic states of baryons, as well as their clustering, at low
to medium redshift $z<5$, has been studied, using both numerical as
well as analytical techniques. In their studies, \cite{VSS02} has
used the hierarchical ansatz, to study the {\em phase-diagrams} of
cosmological baryons as function of redshift. The low temperature
``cool'' component of the intergalactic medium (IGM) represented by
Lyman-$\alpha$ forest typically satisfies the constrain $10^3{\rm
K}<{\rm T}<10^4{\rm K}$. The exact values of the lower and
upper-limit depends somewhat on the redshift. The ``cold'' component
of the IGM is very well characterized by a well-defined equation of
state. The ``warm'' component of the IGM on the other hand is
shock-heated to a temperature range of $10^4{\rm K}<{\rm T}<10^7{\rm
K}$ due to the collapse of non-linear structure and can not be
defined by a well defined equation of state. Though the ``warm''
component does follow a mean temperature-density relation, the
scatter around this relation however is more significant than for
the ``cool'' component. Both the ``cool'' and ``warm'' components
originate outside the collapsed halos and typically reside in
moderate overdensites $1+\delta<100$ \citep{Mu11a}. Finally the remaining
contribution comes from the hot baryonic component of the virialized
high density halos with temperatures ${\rm T}>10^7{\rm K}$.

In  Figure \ref{fig:skew} and  Figure \ref{fig:kurt} we have plotted the skewness and kurtosis parameters
from $y_{\rho}$ maps made using only the medium over-density regions $1+\delta<100$ 
which is caused by both ``warm'' and the ``cool'' component of the
IGM, as well as the map made from low temperature component $y_{\rm T}$.
The contribution from high density virialized halos are
shown too. The skewness and kurtosis corresponding to these individual
components are computed individually. The results are presented for the same
range of smoothing scales considered before.
It is interesting to note that for GO simulations the skewness computed using component maps
$y_\rho$ and $y_{\rm T}$  too are very close to predictions from lognormal model though
clearly due to truncation of high $\rho$ tail of the underlying density PDF responsible
for construction of $y_\rho$ maps the $S_N$ parameters computed from $y_\rho$ maps are
lower than their counterparts constructed from $y$ maps.
In PC simulations only $y_{\rm T}$ maps agree with lognormal model as explained above the low temperature
regions are less likely to be disturbed by non-gravitational processes. This is
in agreement with our previous finding for the entire PDF \citep{Mu11a}. 
The cumulant correlators for y-maps i.e. $C_{21}$ and 
mixed cumulant correlators $C^{y\kappa}_{21}$ follow a similar patterns. 

\begin{figure}
\begin{center}
{\epsfxsize=15.5 cm \epsfysize=5. cm {\epsfbox[21 525 587 712]{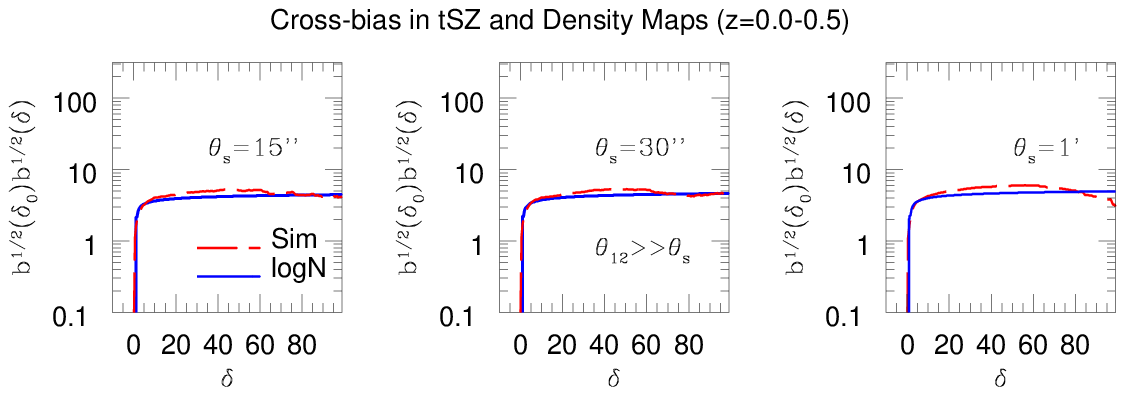}}}
\end{center}
\caption{The cross-bias function  ${\cal B}(>\delta y, >\kappa)$ defined in Eq.(\ref{eq:def_b}) is plotted as a function of $\delta y/\la y\ra$.
The solid lines show the predictions from 
the lognormal distribution. The simulation (GO) results are plotted with long-dashed lines
Three different smoothing
angular scales are considered $\theta_s=\cs{15}''$ (left-panel), $\theta_s=\cs{30}''$ (middle-panel) and $\theta_s=\cs{1}'$ (right-panel) 
respectively. The density threshold was fixed at it saturation value and the cross-bias was computed as a function of $y$-threshold. 
\cs{The redshift bin considered is $z=0.0-0.5$}.}
\label{fig:mbias1}
\end{figure}

\begin{figure}
\begin{center}
{\epsfxsize=15.5 cm \epsfysize=5. cm {\epsfbox[21 525 587 722]{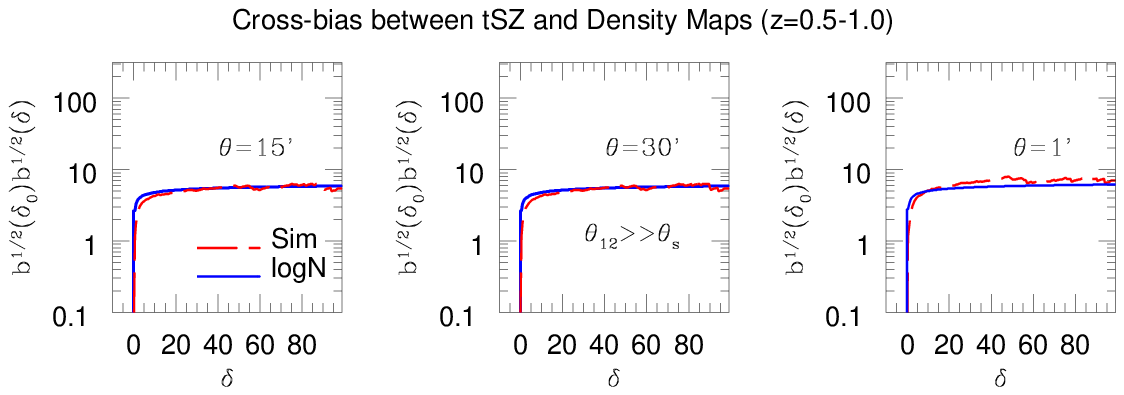}}}
\end{center}
\caption{The mixed bias ${\cal B}(>\delta y,>\kappa )$ defined in Eq.(\ref{eq:def_b}) is plotted as a function of $\delta y$ for projected mass distribution. Three different smoothing
angular scales are considered $\theta_s=\cs{15}''$ (left-panel), $\theta_s=\cs{30}''$ (middle-panel) and $\theta_s=\cs{1}'$ (right-panel) 
respectively. The simulation results are plotted with long-dashed lines. The log-normal results
are shown with solid lines respectively. The redshift bin considered is $z=0.5-1.0$.
The variance associated with y-maps is higher compared to the $\delta$ maps which explains 
larger dynamic range for $\delta y$.}
\label{fig:mbias2}
\end{figure}

{\bf Integrated Bias:}
The computation of integrated bias $b(>y)$ was carried out using
techniques discussed in \citep{Mu11a} which relies on using a grid
to compute the joint PDF using cells separated by a distance. Indeed,
the computation of joint PDF is equivalent to a computation of the bias function
$b(\delta)$. The differential bias defined this way is noisy so
we use the integrated bias $b(>\delta_0)$ beyond a fixed threshold $\delta_0$
as a stable indicator (see Eq.(\ref{eq:joint_tomography})). The results for our computation for bias
are presented in Figure \ref{fig:bias1} and Figure \ref{fig:bias2}.
The bias function carry information regarding cumulant correlators to all orders.
 We find that the analytical predictions for bias to match accurately with the
one recovered from numerical simulations. The bias associated with
joint PDF of $\kappa$ and $y$ parameters are defined in Eq.(\ref{eq:joint_pdf}). These
bias functions are related to mixed cumulant correlators defined in Eq.(\ref{eq:def_b}).
These results are presented in Figure \ref{fig:mbias1} and Figure \ref{fig:mbias2}. 
The recovery of mixed bias function follow exactly the similar procedure
and involve cross-correlating cells from two different maps at a fixed distance. 
We used the analytical predictions for the individual bias function for $\kappa$ and $y$ 
maps to compute the {\em mixed} bias function, which is essentially related to
the joint PDF of both $\kappa$ and $y$ maps to cross specific thresholds (not necessarily the same).
The results again match with  lognormal predictions. The mixed bias function contains the knowledge
of mixed cumulants of all order.  Three different smoothing angular scales were considered
for computation of the integrated bias $\theta_s= 15'',30'',1'$. The bias is weakly dependent
on $\theta_s$.

To summarize our results about PC simulations, the smoothness
of these maps are reflected in their low variance. This is
primarily due to the high level of the pre-heating that erases many
substructures resulting in maps with less features. We include these
simulations in our studies mainly to test the limitations of
analytical predictions. The fundamental assumption in our analytical
modeling is of gravity induced structure formation where baryons are
considered as the biased tracers of underlying dark matter
clustering. We find significant deviation of the numerical results
from theoretical predictions in the presence of high level
preheating at small angular scales (see Figure \ref{fig:tsz_nc}).
These deviations are more pronounced at smaller angular scales. The
PDFs become Gaussian at scales $\theta_s \sim 10'$ or larger. The
deviation at all-scales is less pronounced if we remove the
collapsed objects and focus on the maps with overdensites $1+\delta
< 100$. However the PDF of the $y$ distribution from ``cold''
intergalactic gas even in PC simulations is represented very accurately by our analytical
results at all angular scales in the presence of pre-heating.
%
%
%
%
\section{Conclusion}
\label{sec:conclu}
Extending our previous work \citep{Mu11a}, we have presented a detailed analysis of the higher-order (non-Gaussian) cross-correlation of
tSZ maps and the projected or redshift-resolved tomographic maps from weak lensing surveys.
We use state-of-the-art {\em Millennium Gas Simulations} that includes realistic description of
gas physics test various ingredients in our theoretical calculations and find reasonable agreement.
Below we list our main results.

{\bf Tomographic resolution and evolution with redshift:}
What distinguishes our work in this paper is that it extends earlier studies which have focused mainly on cross-correlation analysis at the level of power spectrum.
The tSZ maps as well as the maps derived from weak lensing surveys are intrinsically non-Gaussian
even at large angular scales. In this study we extend the cross-correlation statistics to higher-order.
We employ the mixed cumulants and their correlators that can extract higher-order statistics
that are useful in probing the thermal history
of the Universe when studied in association with redshift resolved tomographic weak lensing data.
The statistics that we introduce can be used in projection (2D)
or with tomographic information, independent of any analytical modelling
to analyse simulations or observational data.
The estimators presented here are generic and can be used for
other cross-correlation studies involving two different {\em arbitrary} fields e.g. kSZ and lensing or soft X-ray and tSZ.
The mixed cumulants and their correlators defined in this paper are the lower order moments of the corresponding one-point or
two-point PDFs and associated bias. These statistics generalize cumulant correlators
defined in \citep{Mu00} for weak lensing surveys and in \citep{Mu11a,Mu11b} for tSZ datasets, and are
useful in cross-correlating these two data sets.

{\bf Beyond order-by-order calculation:}
Beyond the order-by-order calculation, a generic prediction for the cumulants or their correlators to an
arbitrary order is required. This was achieved by adopting the generating function formalism inherent
in models based on the {\em hierarchical ansatz}. We have employed a generic version of the hierarchical ansatz
as well as the lognormal distribution to model the underlying mass distribution.
Both of these have been studied extensively in the literature.
These particular models go beyond the lowest order in correlation hierarchy.
For the hierarchical model, the generating function is parametrized
in terms of a single parameter $\omega$ which is fixed using inputs from numerical simulations, and
is the only freedom allowed in this model. In the perturbative regime
this parameter can be calculated analytically. Indeed, modelling
of pressure fluctuations in terms of the underlying density distribution is more
complicated, and we rely on a redshift dependent linear biasing model
that is expected to be valid at large angular scales~\citep{Mu11a,Mu11b}.
More complicated models can be considered in this framework using additional inputs
from numerical simulations; here we have tried to keep the modelling as simple as possible.
The halo model can be used for the construction of
the lowest order cumulant correlators. However, in such an approach, results can be derived only in an
order-by-order manner and
the relevant PDF and bias needs to be constructed using suitably
truncated Edgeworth expansions. As mentioned previously, the PDF and bias for the the weak lensing
data have been studied by many authors using both the hierarchical ansatz and the
lognormal distribution. The studies have shown how accurately these distributions
can model numerical simulations. In more recent studies, the hierarchical ansatz has
also been used successfully in tSZ modelling.
Based on this success, our primary goal in this study has been to extend such analysis to
compute the joint PDF of $\kappa$ and $y$ along same or different lines of sight. In lieu of individual haloes, we consider
here the diffuse tSZ component that correlates with the large-scale weak lensing shear and convergence.

{\bf New variables to isolate background cosmology:}
We have shown that the two different variables, i.e. the scaled tSZ parameter $\eta' = \delta y/\la y\ra$ and the reduced
convergence field $\eta= \kappa/\kappa_m$ can be introduced to simplify the analysis. Employing the hierarchical ansatz, we next show
that the statistics of these reduced variables are the same as
the underlying density $\rho/\rho_b=1+\delta$ under certain simplifying approximations, which is true both at the level of one and two-point statistics.
Thus, the final results do not depend on the specific details of the input parameters of hierarchical ansatz, and may point to a
wider applicability. The joint distribution involving $\kappa$ and $y$ can thereby be
written in terms of $p(\eta)$ and $b(\eta)$ with suitable scaling involving $\la y \ra$ and $\kappa_m$.
The detailed construction of the PDF is insensitive to modelling of the bias of the
hot ionized gas with respect to the dark matter distribution. The detailed prescription
for such bias enters the results through the computation of the variance and the two-point
correlation function $\xi_{y\kappa}(\theta_{12})$.

{\bf Linear biasing:}
Specific models of hierarchical clustering have previously been employed to understand the
tSZ effect (see e.g. \citep{VJS01,VS99,VS00}). However, they have been employed to model contribution from the collapsed objects (e.g. massive clusters).
In this case, the tail of the scaling functions $h(x)$ and $b(x)$ are
relevant as they describe the collapsed objects.
In addition to these inputs, the modelling would involve a prescription
for the hydrodynamic equilibrium of the gas residing in the halo.
These can be used to model the PDF of $y(\oh)$, i.e. $p_y(y)$ or $b_y(y)$.
The modelling considered here is complementary to
our analysis. We have considered redshift dependent linear biasing to model the
shock heated gas in the inter-galactic medium that produces a diffuse
tSZ effect. The linear biasing model has been studied at the level of the
power spectrum by many authors \citep{GS99a,GS99b, PS00}.
The hierarchical model that we use here was also previously used by \cite{ZPW02},
\cite{CTH00} and \citep{AC1,AC2} for modelling lower order moments where the linear
biasing model for the pressure fluctuations was assumed.
Our results are natural generalizations of such studies to higher order multi-spectra
to study the non-Gaussian aspects of tSZ maps.

{\bf Effect of Pre-heating:}
It is known that the IGM is most likely have been preheated by non-gravitational
sources. The feedback from SN or AGN can play an important role.
The analytical modelling of such non-gravitational processes is rather
difficult. Numerical simulations \citep{Spr01,dasilva00,dasilva04,White02,Lin04} have shown that the amplitude of the tSZ
signal is sensitive to the non-gravitational processes, e.g. the amount of radiative cooling and energy feedback.
It is also not straightforward to disentangle contributions from
competing processes. The inputs from simulations are vital for any progress.
Our analytical results should be treated as
a first step in this direction. We have focused mainly on large angular scales
where we expect the gravitational process to dominate and such effects to
be minimal. We find that the simulations without pre-heating (GO) can be understood
using analytical arguments. Thus the affect of additional baryonic physics
can be separated using the formalism developed here. To understand the effect
of baryonic physics we have developed the simulated maps in different components
and studied them individually. We also find that even in the presence of 
preheating the {\em cool} component $y_{\rm T}$ of the IGM remains relatively 
undisturbed and can be modelled using a linear biasing scheme. 

{\bf From tomography to 3D:}
Our results are derived using cross-correlating tomographic slices from weak lensing surveys
and the projected tSZ surveys. With suitable modification, they can be equally applicable to cross-correlation studies
using tomographic maps from the same or different weak lensing surveys. The results can also be used to study
the weak lensing of CMB by cross-correlating it against tomographic weak lensing maps or tSZ maps.
To move beyond a tomographic or projected survey, a complete 3D analysis can be invoked.
An analysis similar to what has been presented here for the kinetic Sunyaev-Zel'dovich effect (e.g. \cite{Shao11})
can provide valuable information about the reionization history of the Universe (Munshi et al. 2013; in preparation).
%


\section{Acknowledgements}
\label{acknow}
DM and PC acknowledges support
from STFC standard grant ST/G002231/1 at School of Physics and
Astronomy at Cardiff University where this work was completed.
SJ and JS acknowledge support from the US Department of Education through
GAANN at UCI.
We thank Alan Heavens for useful discussions. It is also a pleasure for us to thank
Martin Kilbinger for related collaboration. We
thank Francis Bernardeau for supplying a copy of his code
which was used to compute the bias and the PDF from hierarchical ansatz.
\bibliography{paper.bbl}
\appendix
\section{Gravitational Clustering  and the Hierarchical Ansatz}
The lower order cumulants and cumulant correlators as well as the one- and two-pint PDFs
are commonly used statistics in cosmology to qunatify clustering \citep{Bernardreview02}.
Use of generating function techniques to go beyond order by order calculations
is an important milestone in this direction \citep{Peebles71, BS89,B92, BS92,B94}. This approach is
complimentary to the halo model based approach that rely on Press-Sechter mass formalism \citep{PS74}. We provide a
very brief review of this formalism in this appendix.
\subsection{The Generating Function}
In scaling analysis of the probability distribution function (PDF) the
void probability distribution function (VPF), denoted as $P_v(0)$, plays most fundamental
role, which can related to the generating function of the cumulants
or $S_N$ parameters, $\phi(z)$ \citep{Wh79,BS89}:
\begin{equation}
P_v(0) = \exp ( -\bar N \sigma(N_c) ) = \exp \Big ( - { \phi (N_c)/
\bar \xi_2^{\delta}} \Big  ).
\end{equation}
\n
Where $P_v(0)$ is the probability of having no ``particles'' in a cell of
of volume $v$, $\bar N$ is the average occupancy of these ``particles'' and
$N_c = \bar N {\bar \xi^{\delta}}_2$ and $\sigma(z) = -{\phi(z)/ z}$. Here
$\phi(z) = \sum_{p=1}^{\infty} { S_p/p! }\; z^p$ is a generating function for $S_N$ parameters.
The genrating function ${\cal G}(\tau)$ for the vertex amplitudes $\nu_p$
is related to $\phi(z)$.
A more specific model for ${\cal G}(\tau)$ is  generally used
to make more specific predictions \citep{BS89}:
\begin{equation}
{\cal G}_{\delta}(\tau) = \Big ( 1 + {\tau/\kappa_a} \Big )^{-\kappa_a}.
\end{equation}
\n
The parameter $\kappa_a$ is related to the parameter $\omega_a$
to be introduced later. The two generating function are related
by the following expression \citep{BS89,BS92,BS99}:
\be
\phi(z) = z {\cal G}_{\delta}(\tau) - { 1 \over 2} z {\tau} { d \over d
\tau} {\cal G}_{\delta}(\tau); \quad \tau = -z { d \over d\tau} {\cal G}_{\delta}(\tau).
\ee
However a more detailed analysis is needed to include the effect of
correlation between two or more correlated volume element which will
provide information about bias, cumulants and cumulant correlators
\citep{BS92, MCM99a,MCM99b,MMC99c}
Notice that $\tau(z)$ (a bias also denoted by $\beta(z)$ in the literature)
plays the role of generating function for
factorized cumulant correlators $C^{\eta\eta'}_{p1}$ ($C_{pq} = C^{\eta\eta}_{p1}C^{\eta'\eta'}_{q1}$):
$\tau(z) = \sum_{p=1}^{\infty} {C^{\eta\eta}_{p1}/p!}\;z^p$.
\subsection{Generating Functions and the Construction of the PDF and Bias}
The hierarchical form of higher order correlation functions appear in two completely
different regimes in gravitational clustering. The generating function approach
is convenient technique to sum up to arbitrary information that leads to construction
of entire probability distribution functions.
\subsubsection {The Highly Non-linear Regime}
The PDF $p(\delta)$ and bias $b(\delta)$  can be related to their
generating functions VPF $\phi(z)$ and $\tau(z)$ respectively
by following equations \citep{BS89,BS92,BS99}
\be
p(\delta) = \int_{-i\infty}^{i\infty} { dz \over 2 \pi i} \exp \Big [ {(
1 + \delta )z - \phi(z)  \over \bar \xi_2} \Big ]; \quad
b(\delta) p(\delta) = \int_{-i\infty}^{i\infty} { dz \over 2 \pi i} \tau(z) \exp \Big [ {(
1 + \delta )z - \phi(z)  \over \bar \xi_2} \Big ] \label{ber}.
\ee
\n
It is clear that the function $\phi(z)$ completely determines
the behaviour of the PDF $p(\delta)$ for all values of $\delta$. However
different asymptotic expressions of $\phi(z)$ govern the behaviour
of $p(\delta)$ for different intervals of $\delta$. For large $y$ we
can express $\phi(z)$ as: $\phi(z) = a z^{ 1 - \omega}$.
Here we have introduced a new parameter $\omega$ for the description of
VPF.
Typically initial power spectrum with spectral index $n=
-2$ (which should model CDM like spectra we considered in our
simulations at small length scales) produces a value
of $.3$ which we will be using in our analysis \citep{CBH96,CBS96,CBBH97}.
The VPF $\phi(z)$ and its two-point analogue $\tau(z)$
both exhibit singularity for small but negative value of $z_*$,
\be
\phi(z) = \phi_* - a_* \Gamma(\omega_*) ( z - z_*)^{-\omega_*}, ; \quad
\tau(z) = \tau_* - b_* ( z - z_* )^{-\omega_* - 1}.
\ee
\n
For the factorizable model of hierarchical clustering the
parameter $\omega_*$
takes the value $-3/2$ and $a_*$ and $b_*$ can be expressed in terms
of the  nature of the generating function ${\cal G}_{\delta}(\tau)$ and its
derivatives near the singularity $\tau_*$
\citep{BS92}:
\be
a_* = {1 \over \Gamma(-1/2)}{\cal G}'_{\delta}(\tau_*) {\cal G}''_{\delta}(\tau_*) \left [
{ 2 {\cal G}'_{\delta}(\tau_*) {\cal G}''_{\delta}(\tau_*) \over {\cal G}'''_{\delta}(\tau_*)}
\right ]^{3/2}; \quad
b_* = \left [
{ 2 {\cal G}'_\delta(\tau_s) {\cal G}''_\delta(\tau_*) \over {\cal G}'''_\delta(\tau_*)}
\right ]^{1/2}.
\ee
As mentioned before the parameter $k_a$ which we have introduced in
the definition of
${\cal G}_{\delta}(\tau)$ can be related to the parameters $a$ and $\omega$ appearing
in the asymptotic expressions of $\phi(z)$ \citep{BS89,BS92}
\be
\omega = k_a / ( k_a + 2),\label{ka}; \quad
a = {k_a + 2 \over 2} k_a^{ k_a /  k_a + 2}.
\ee
Similarly the parameter $z_s$ which describe the behaviour
of the function $\phi(z)$ near its singularity can be
related to the behaviour of
${\cal G(\tau)}$ near $\tau_s$ which is the solution of the equation
\citep{BS89,BS92}
\begin{equation}
\tau_* = {{\cal G}'_{\delta}(\tau_*)/{\cal G}''_{\delta}(\tau_*) },
\end{equation}
\n
finally we can relate $k_a$ to $y_*$ by following expression (see Eq.~(\ref{ka})):
\be
z_* = - { \tau_* \over {\cal G}'(\tau_*)}; \quad
-{ 1 \over z_*} = x_{\star} = {1 \over k_a } { (k_a + 2)^{k_a + 2} \over (k_a + 1)^{k_a+1}}.
\ee
\n
The newly introduced variable $x_\star$ will be useful to define the
large-$\delta$ tail of the PDF $p(\delta)$ and the bias $b(\delta)$.
Different asymptotes in $\phi(z)$
are linked with behaviour of $p(\delta)$ for various regimes of
$\delta$. For very large values of variance i.e. $\xi_2$
it is possible to define a scaling function $p(\delta) = {
} h(x)/\bar \xi_2^2 $  which will encode
the scaling behaviour of PDF, where $x$ plays the role of the scaling
variable and is defined as $x={(1 + \delta)}/ \bar \xi_2$. We list
different ranges of $\delta$ and specify the behaviour of $p(\delta)$
and $b(\delta)$ in these regimes \citep{BS89}
\begin{equation}
{(\bar \xi_2^{\delta}) }^{ - \omega/( 1 - \omega)} << 1 + \delta << \bar \xi_2^{\delta};
~~~~~~
p(\delta) = { a \over (\bar\xi_2^{\delta})^2} { 1- \omega \over \Gamma(\omega)}
\Big ( { 1 + \delta \over \bar \xi_2^{\delta} } \Big )^{\omega - 2}; ~~~~~
 b(\delta) = \left ( {\omega \over 2a } \right )^{1/2} { \Gamma
(\omega) \over \Gamma [ { 1\over 2} ( 1 + \omega ) ] } \left( { 1 +
\delta \over \bar \xi_2^{\delta}} \right)^{(1 - \omega)/2}
\end{equation}
\begin{equation}
1+ \delta >> {\bar \xi_2^{\delta}}; ~~~~
p(\delta) = { a_s \over (\bar\xi_2^{\delta})^2 } \Big ( { 1 + \delta \over \bar
\xi_2^{\delta}}  \Big ) \exp \Big ( - { 1 + \delta \over x_{\star} \bar \xi_2^{\delta}}
\Big );  ~~~~~ b(\delta) = -{ 1 \over {\cal G}'(\tau_s)} {(1 + \delta)
\over { {\bar \xi_2^{\delta}}}}
\end{equation}
\n
The integral constraints satisfied by scaling function are
$ S_1 = \int_0^{\infty} x h(x) dx = 1$ and
$ S_2 = \int_0^{\infty} x^2 h(x) dx = 1$. These take care of
normalization of the function $p(\delta)$. Similarly the
normalization constraint over $b(\delta)$ can be expressed as
$C_{11} = \int_0^{\infty} x b(x)h(x)dx = 1$, which translates into
$\int_{-1}^{\infty} d\delta b(\delta)p(\delta) = 0$ and
$\int_{-1}^{\infty} d\delta \delta b(\delta)p(\delta) = 1$.
Several numerical
studies have been conducted to study the behaviour of $h(x)$ and $b(x)$
for different initial conditions see e.g. \citep{CBH96,CBS96,CBBH97}. For very small values of $\delta$ the behaviour of
$p(\delta)$ is determined by the asymptotic behaviour of $\phi(z)$
for large values of $y$, and it is possible to define another scaling function
$g(z)$ which is completely determined by
$\omega$, the scaling parameter can be expressed as $z' = (1+
\delta)a^{-1/(1-\omega)}({\bar \xi^{\delta}}_2)^{\omega /(1 - \omega)}$.
However, numerically it is much simpler to determine $\omega$
from the study of $\sigma(y)$ compared to the study of $g(z)$.
\begin{equation}
1 + \delta << \bar \xi_2^{\delta};~~~~
p(\delta) = a^{ -1/(1 - \omega)} ({\bar \xi^{\delta}}_2)^{ \omega/(1 -
\omega)} \sqrt { ( 1 - \omega )^{ 1/\omega } \over 2 \pi \omega z'^{(1
+ \omega)/ \omega } } \exp \Big [ - \omega \Big ( {z' \over 1 - \omega}
\Big )^{- {{(1 - \omega)}/\omega}} \Big ]; ~~~~~~~b(\delta) = -
\left ( {2 \omega \over \bar{ \xi}^{\delta}_2} \right )^{1/2} \left ({ 1 -
\omega \over z'}  \right )^{(1 - \omega)/2 \omega}
\end{equation}
\n
To summarize, we can say that the entire behaviour of the PDF
 $p(\delta)$ is
encoded in two different scaling functions, $h(x)$ and $g(z')$ and one can
also study the scaling properties of $b(\delta)$ in terms of the scaling
variables $x$ and $z$ in a very similar way.
%
\subsubsection{The Quasi-linear Regime}
The first departure from Gaussianity can be studied analytically
using perturbative techniques \citep{B92,B94} as well as using
steepest descent methods \citep{V00a,V0b,V02}. These techniques are
also invaluable in modelling the statistics of extreme underdensities.
The PDF and bias now can be expressed in terms of $G_{\delta}(\tau)$ \citep{B92,B94}:
\begin{eqnarray}
&&p(\delta)d \delta = { 1 \over -{\cal G}_{\delta}'(\tau) } \Big [ { 1 - \tau {\cal G}_{\delta}''(\tau)
/{\cal G}_{\delta}'(\tau) \over 2 \pi {\bar \xi^{\delta}}_{2} }  \Big ]^{1/2} \exp \Big ( -{ \tau^2
\over 2 {\bar \xi^{\delta}}_{2}} \Big ) d \tau; ~~~~~ b(\delta) = - \left (
{k_a \over \bar \xi^{\delta}_2} \right ) \left [ ( 1 + {\cal G}_{\delta}(\tau)
)^{1/k_a} - 1 \right ] , \\
&&{\cal G}_{\delta}(\tau) = {\cal G}(\tau) - 1 =  \delta.
\end{eqnarray}
\n
The above expression is valid for $\delta < \delta_c$ where the $\delta_c$
is the value of $\delta$ which cancels the numerator of the pre-factor
of the exponential function appearing in the above expression. For
$\delta > \delta_c$ the PDF develops an exponential tail which is
related to the presence of singularity in $\phi(z)$ in a very similar
way as in the case of its highly non-linear counterpart \citep{B92,B94}:
\begin{equation}
p(\delta) d \delta = { 3 a_s \sqrt {{\bar \xi^{\delta}}_2} \over 4  {\sqrt \pi} }
\delta^{-5/2} \exp \Big [ -|z_s|{ \delta \over {\bar \xi^{\delta}}_{2}} + {|\phi_s|
\over {\bar \xi^{\delta}}_{2}} \Big ] d \delta; ~~~~~b(\delta) = -{ 1 \over
{\cal G}'(\tau_s)} {(1 + \delta)
\over { {\bar \xi^{\delta}}_2}}.
\end{equation}
The quasilinear regime remains relatively better understood than the highly nonlinear
regime, as the connection to dynamics can be made using analytical schemes. Attempts
have been made to extend the perturbative results to the highly nonlinear regime \cite{CBBH97}.
The lower order cumulant correlators are related to the moments of the bias function.
For the case of top-hat smoothing and spectral index $n_e$ the lowest order
of the cumulant correlator is given by \citep{B94}:  $C^{\eta\eta'}_{21} = {68/21}-(n_e+3)/3$.
\section{The Lognormal Distribution}
In addition to the hierarchical ansatz {\em lognormal} distribution too is a popular analytical
model commonly used in cosmology \citep{Ham85,CJ91,Bouchet93,Kf94,BK95,Col94}. It appears in the quasi-linear
regime \citep{MSS94} as a natural outcome of gravitational dynamics, under certain simplifying assumptions \citep{MLMS92}.
Lognormal distribution is routinely used to model
the statistics of weak lensing observables \citep{Mu00,TTHF02}, clustering of Lyman-$\alpha$
absorption systems (e.g. \cite{BD97}) and more recently by \citep{Mu11a} for modelling of the tSZ statistics.

To understand the construction of lognormal distribution, we introduce a Gaussian PDF in variable $s$:
$ p(s) = (2 \pi \bar\Xi)^{-1/2} \exp[-(s-\mu)^2/2\bar\Xi] $. With a change of variable $s = \ln(r)$ we can
write down the PDF of $y$ which is a lognormal distribution $p(r) = (2\pi\bar\Xi)^{-1/2} \exp [-(\ln(r) -\mu)^2/2\bar\Xi]/r$.
The extra factor of $(1/r)$ stems from the fact: $dr/r = ds$. Note that $s$ is positive definite and
is often associate with $\rho/\la\rho\ra = 1 +\delta$ which means $\la r \ra =1$.
The moment generating function for the lognormal in terms of the mean $\mu$ and the variance $\bar\Xi$ has the
following form: $\la r^n \ra = \exp(n\mu + n^2 \bar\Xi/2)$. This however leads to the fact
that if the underlying distribution of $r$ or the density is Gaussian we will have to impose the
condition: $\mu=-\bar\Xi/2. \label{eq:consistency}$.
In our notation above $\Xi$ is the distribution of the underlying
Gaussian field. The variance of $r$ defined as $\la r^2 \ra - \la r \ra^2 = \exp(\Xi) -1 = \xi_2^{\delta}$. So we can write
$\bar\Xi = \ln(1+\bar\xi_2^{\delta})$. This is the result that was used above.
The generalisation to two-point or bivariate PDF can be done following the same arguments and can be found in \citep{KTS01}.

In the limit of large separation $\Xi_{12}\rightarrow 0$ we can write down the two point PDF
\be
p_{\rm ln}(\delta_1,\delta_2)= p_{\rm ln}(\delta_1)p_{\rm ln}(\delta_2)[1+ b_{\rm ln}(\delta_1)\xi^{\delta}_{12}b_{\rm ln}(\delta_2)];
\quad\quad  b_{\rm ln}(\delta_i)= \Lambda_i/\bar\Xi_{}.
\ee
However, it is simpler to estimate the cumulative or integrated bias associated with objects beyond a certain density threshold $\delta_0$.
This is defined as $ b(\delta>\delta_0)=\int_{\delta_0}^{\infty} p(\delta) b(\delta) d\delta / \int_{\delta_0}^{\infty} p(\delta) d\delta$.
In the low variance limit $\bar \xi_2^{\delta} \rightarrow 0$ the usual Gaussian result is
restored $b(\delta)= \delta/\bar \xi_2^{\delta}$. The parameters $\Lambda,\Lambda_i, \Xi_{12}, \bar\Xi$ that
we have introduced above can be expressed in terms of the two-point (non-linear) correlation function
$\xi^{\delta}_{12} = \la \delta_1\delta_2 \ra$ and $\bar \xi^{\delta}_2$ is the volume average of the non-linear two-point correlation
function $\xi_{12}$ of the smoothed density field.

The validity and limitations of various aspects of the one-point and two-point PDFs have been
studied extensively in the literature against N-body simulations \cite{B92,B94}.
It is known that in the weakly nonlinear regime the lognormal distribution is equivalent
to the hierarchical model with a generating function $\cal G(\tau) = \exp(-\tau)$.
This leads to $S_p = p^{p-2}$. The loop level corrections can be computed exactly
for the lognormal distribution, which gives $S_3 = 3 + \bar\xi_2^{\delta}$ and
$S_4 = 16 + 15\bar\xi_2^{\delta} + 6(\bar\xi^{\delta}_2)^2 + (\bar\xi^{\delta}_2)^3$.
It has been shown (see e.g. \cite{KTS01}) that the lognormal distribution very accurately describes the cosmological
distribution functions even in the nonlinear regime $\bar\xi^{\delta}_2 \le 4$ for relatively
high values of density contrast $\delta < 100$.
\end{document}